\documentclass[%
 aps,
 pra,
 reprint,
 superscriptaddress,
]{revtex4-1}
\usepackage{amssymb}
\usepackage{amsmath}
\usepackage{graphicx}% Include figure files
\usepackage{color,xcolor}
\usepackage{bbold}
\usepackage{dcolumn}% Align table columns on decimal point
\usepackage{bm,bbm}% bold math
\usepackage{physics}
\usepackage[colorlinks=true,allcolors=gray!80!blue]{hyperref}% add hypertext capabilities
\begin{document}
%\preprint{APS/123-QED}
\title{Continuous-variable entanglement distillation by cascaded photon replacement}

\author{Yasamin Mardani}
\affiliation{Department of Physics, Isfahan University of Technology, Isfahan 84156-83111, Iran}

\author{Milad Ghadimi}
%\thanks{These two authors contributed equally}
\affiliation{Department of Physics, Isfahan University of Technology, Isfahan 84156-83111, Iran}

\author{Ali Shafiei}
%\thanks{These two authors contributed equally}
%\affiliation{Department of Physics, Isfahan University of Technology, Isfahan 84156-83111, Iran}
\affiliation{Department of Electrical \& Computer Engineering, Isfahan University of Technology, Isfahan 84156-83111, Iran}

\author{Mehdi Abdi}
\affiliation{Department of Physics, Isfahan University of Technology, Isfahan 84156-83111, Iran}

%_______________ABSTRACT_______________
\begin{abstract}
We study the entanglement distillation in continuous variable systems when a photon replacement protocol is employed.
A cascaded protocol is studied and we find that the resultant entanglement increases by increasing the number of repetitions.
Interestingly, the entanglement enhancement is not sensitive to the asymmetry of the protocol and gives the same result for any arrangement in the absence of loss.
The non-Gaussianity of the outcome state is also studied and it is found that the non-Gaussianity of the state dramatically depends on the experimental arrangements.
By providing practical information on photon replacement operation, this work is one step towards realization of universal quantum computation.
In particular, in setups where deGaussifying protocols are only applicable to one of the parties.
\end{abstract}

\maketitle

%≡≡≡≡≡≡≡≡≡≡≡≡≡≡≡SECTION≡≡≡≡≡≡≡≡≡≡≡≡≡≡≡
\section{Introduction \label{sec:intro}}
Most quantum information processing protocols require maximally entangled states or at least high entanglements to guarantee fault-tolerant performance~\cite{Steane1999}.
In real life even after creation of such states the entanglement is bond to decrease due to the environmental effects. To retrieve the entanglement lost into the reservoirs, entanglement distillation protocols must be invoked.
Entanglement distillation refers to the protocols that employ many copies of entangled states to extract a smaller number of states with increased degree of entanglement using local operations and classical communications in quantum systems.
It is usually decomposed into entanglement purification, which is to extract entanglement from mixed states~\cite{Bennett1996} and entanglement concentration, that is achieving a maximally entangled state from pure lower entangled states~\cite{Bennett1996a}.
The concept was first suggested for discrete variable systems, nevertheless, it was later extended to the continuous variable (CV) systems~\cite{Giedke2001}.
An important issue that occurs when it comes about continuous variable concentration is that the Gaussian states cannot be distilled by only using Gaussian operations and classical communications~\cite{Giedke2002, Fiurasek2002, Eisert2002}.
The most known non-Gaussian operations that their vital role in enhancement of entanglement has been proven are photon subtraction~\cite{Opatrny2000, Kitagawa2006, Ourjoumtsev2007, Takahashi2010, Zhang2010}, photon addition~\cite{Zhang2013}, and photon replacement (PR)~\cite{Lvovsky2002}.
These operations distill input Gaussian states into more entangled \textit{non-Gaussian} states.
It is worth mentioning that a scheme has recently been proposed to bypass the necessity of non-Gaussian states/operations by employing assistant parties~\cite{Adesso2019}.

It has been shown that an ideal photon-addition $\hat{a}^{\dagger}$ followed by a subtraction $\hat{a}$ results-in more entanglement than a single ideal photon addition or subtraction when applied to a two-mode squeezed vacuum (TMSV) state~\cite{Yang2009}.
%By the term \textit{ideal} we mean that they used the annihilation operator $\hat{a}$ as photon subtraction, and the creation operator $\hat{a}^{\dagger}$ as the photon addition operations.
A coherent superposition of ideal photon addition and subtraction $c_{1}\hat{a}+c_{2}\hat{a}^{\dagger}$ gives even more entanglement for small original entanglements~\cite{Nha2011}.
Also, a generalized form of such operations is capable of generating entangled coherent states with high degree of entanglement starting from coherent states~\cite{Liu2015}.
One then asks how a cascaded operation of such distillation operations affects the outcome.
The question was thus thoroughly pursued for ideal photon addition and subtraction by Navarrete-Benlloch~\textit{et~al.} in Ref.~\cite{Navarrete2012}. It was shown that the entanglement, as well as the non-Gaussianity of the output state generally increases with the number of operations.
Also, for a given number of operations, the entanglement enhancement resulting from photon addition is greater or equal to that of photon subtraction. In the symmetric case, which is the optimal situation, photon addition and subtraction lead to the same result.

The Gaussian entangled states can also be concentrated into higher entangled non-Gaussian states via single- or two-mode photon replacement operations~\cite{Bartley2015, Xu2015, Hosseinidehaj2015}.
These works have considered the more practical case where the operation is not ideal and occurs with a finite probability.
The comparisons to the photon addition and subtraction cases, show that in the small squeezing regime, the replacement scheme performs better.
And importantly, while the success probabilities of the maximum achievable entanglements in the photon addition and subtraction cases is very low, a photon replacement maximizes the entanglement at reasonable probabilities.
The other advantage of PR is its compatibility with the Gaussification protocols that are widely acknowledged in CV quantum repeaters~\cite{Browne2003, Eisert2004, Campbell2012, Seshadreesan2019}. In these protocols the entanglement is concentrated by a non-Gaussian operation after each swapping and Gaussified before getting swapped into the next level~\cite{Furrer2018}.
Lund and Ralph have used Gaussification protocols after applying PR on noisy Gaussian states and reached more entangled Gaussian states~\cite{Lund2009}.
As a generalization to the PR, multi-photon replacement protocols, where more number of photons are sent and reabsorbed from one or both modes of a TMSV state have also been considered~\cite{Zubairy2017, Birrittella2018}.
The experimental feasibility, however, depends on the performance of photon number resolving detectors with the desired number of photons.

In this paper, we study a cascaded photon replacement (CPR) protocol and its effect on the entanglement distillation of a two-mode squeezed vacuum state and its deGaussification.
This protocol has proven useful for preparation of Gottesman-Kitaev-Preskill states~\cite{Eaton2019}.
By starting from an effective operator for the photon replacement we derive an analytical expression for the output state after a series of cascaded PR operations.
The expression is then used to calculate the success probability of the CPR. We employ logarithmic negativity as a measure of entanglement and provide an analytical expression for it. The results show that the amount of entanglement increases by increasing the number of operations and asymptotically saturates for large number of PR operations.
Furthermore, the result is independent from the mode to which the operation is applied. That is, one may opt to apply CPR in fully symmetric or completely asymmetric fashions as two extremes without any change in the outcome.
The comparison to the cascaded photon addition and subtraction protocols reveals better performance of CPR regarding success probability at their respective maximum distillable entanglement.
The protocol is also practically implementable as it relies on the single-photon states and detectors.
The study of non-Gaussianity shows an interesting behavior. Indeed, the non-Gaussian nature of the outgoing state can be controlled by engineering the setup properties.

The rest of paper is organized as follows:
In Sec.~\ref{sec:pr} we provide an effective operator for a single one-mode photon replacement. The operator is used in Sec.~\ref{sec:cpr} to study the cascaded photon replacement on TMSV states and the entanglement properties of the outcome state.
In Sec.~\ref{sec:result} the entanglement and non-Gaussianity properties of the protocol are studied by numerical evaluations and the resuts are compared to the cascaded PA and PS cases.
The paper is concluded in Sec.~\ref{sec:concl}.

%≡≡≡≡≡≡≡≡≡≡≡≡≡≡≡SECTION≡≡≡≡≡≡≡≡≡≡≡≡≡≡≡
\section{Photon replacement \label{sec:pr}}
\begin{figure}[t]
\includegraphics[width=\columnwidth]{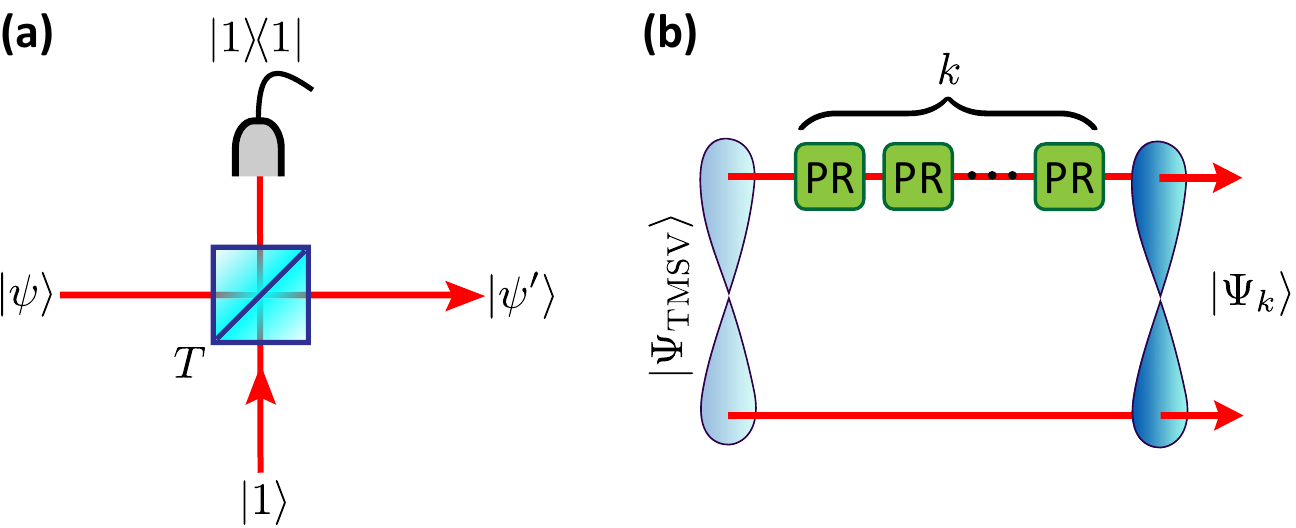}
\caption{\label{structure}%
(a) A photon replacement operation: One part of a bi- or multi-partite state, $\ket{\psi}$ is mixed with a single-photon state $\ket{1}$ via a beam-splitter with transmissivity $T$. One of the beam-splitter outputs is measured by a single-photon detector. The other output is let to pass provided arrival of a single-photon is registered by the detector.
(b) The fully asymmetric CPR protocol applied to a bipartite TMSV entangled state. Each green box incorporates a PR operation that includes: a single-photon source, a beam-splitter, and a single-photon detector. The output of a PR operation is set as the input of the next one.}
\end{figure}

We first review the physics and formulation of the photon replacement protocol, also called photon catalysis or quantum-optical catalysis.
In the PR scheme, the input mode of one or all parties are mixed with a single-photon state through a beam-splitter and one of the output ports of the beam-splitter is measured by a single-photon detector (SPD).
Once the single-photon detector is triggered (a single-photon is detected), the other output port is allowed to pass [Fig.~\ref{structure}(a)].
%This protocol is called photon replacement because it is like the single-photon detected at the output getting replaced by the single-photon that has been sent through the other input port.
The setup of photon addition and subtraction protocols are very similar to that of PR. However, the differences are in the input of the ancillary mode of the beam-splitter or the conditional measurement at the output ports.
In the photon subtraction the beam-splitter port remains free (vacuum input), while in the photon addition the output is post-selected conditioned on no detection in the SPD.
%~~~~~~~~~~~~~~~FIGURE~~~~~~~~~~~~~~~
%~~~~~~~~~~~~~~~~~~~~~~~~~~~~~~~~~~~~

The effect of a beam-splitter with transmissivity $T$ on pure input modes can be described by the operator $\hat{B}^{\dagger}$ such that~\cite{Leonhardt1997}
\begin{equation*}
\hat{B}^{\dagger}\ket{n_{1},n_{2}}=\hspace{-2mm}\sum_{k_{1},k_{2}=0}^{n_{1},n_{2}}\hspace{-2mm}b_{k_{1},k_{2}}^{n_{1},n_{2}} \ket{k_{1}+k_{2}, n_{1}+n_{2}-k_{1}-k_{2}},
\label{BS}
\end{equation*}
where the coefficients $b_{k_{1},k_{2}}^{n_{1},n_{2}}$ are
\begin{align}
b_{k_{1},k_{2}}^{n_{1},n_{2}}&=\frac{1}{\sqrt{n_{1}\!!\ n_{2}\!!}} \binom{n_{1}}{k_{1}} \binom{n_{2}}{k_{2}} T^{n_{2}+k_{1}-k_{2}}R^{k_{2}}(-R)^{n_{1}-k_{1}}  \nonumber\\
&~~~\times\sqrt{(k_{1}+k_{2})!(n_{1}+n_{2}-k_{1}-k_{2})!}~.
\end{align}
The ideal detection of a single-photon at the detector is equivalent to applying the projection operator $|1\rangle\!\langle1|$ on the corresponding subsystem. Hence, applying the PR protocol on a state $\ket{\psi}$, is formulated by
\begin{equation}
\ket{\psi^{\prime}}_1\!\ket{1}_2=(\mathbbm{1}_1\otimes|1\rangle\!\langle1|_2)\hat{B}^{\dagger}\ket{\psi}_1\!\ket{1}_2,
\label{PR_Calculation}
\end{equation}
where $\mathbbm{1}$ represents the identity operator and does not change the state of the corresponding subsystem.
By employing Eq.~(\ref{PR_Calculation}), one arrives at the following effective operator that describes the effect of photon replacement
\begin{equation}
\hat{R}=\sum_{n=0}^{\infty}T^{n-1}\left[T^{2}-n(1-T^{2})\right] |n\rangle\!\langle n|,
\label{effect-operator}
\end{equation}
where $T$ is the transmissivity of the beam-splitter.

\section{Cascaded photon replacement \label{sec:cpr}}
The CPR protocol is achieved by performing a sequence of photon replacement operations on each or a few of the parties.
Fig.~\ref{structure}(b) shows the setup of a CPR protocol symmetrically applied to a bipartite entangled state.
Here, we put our focus to the two-mode-squeezed-vacuum states.
In Schmidt form, these states are given by
\begin{equation}
	\ket{\Psi_{\!\textsc{tmsv}}}=\sqrt{1-\lambda^2} \sum_{n=0}^{\infty} \lambda^{n} \ket{n,n},
\label{TMSV}
\end{equation}  
where we have adapted the simple notation $\ket{n,n}\equiv\ket{n}_1\!\ket{n}_2$ and $\lambda=\tanh(r)$ and $r$ is the squeezing parameter. TMSV states are Gaussian, entangled, and physical for $0<\lambda<1$.
From the photon replacement effective operator in \eqref{effect-operator} one easily finds that for both symmetric and asymmetric cases, the resulting states are of the following form
\begin{equation}
\ket{\Psi}=\mathcal{N}\sum_{n} c_{n} \ket{n,n},
\label{general_form}
\end{equation}
where the expansion coefficients $c_n$ should be determined and are such that $\mathcal{N} = \big(\sum_{n} c_{n}^2\big)^{-1/2}$, hence, the state $\ket{\Psi}$ remains normalized.
%Note that the TMSV state (\ref{TMSV}) is normalized.
For calculating the outcome of each PR step, one applies the effective operator (\ref{effect-operator}) and normalizes the resulting state. Then the sum over coefficients of the un-normalized state $P=\sum_{n} c_{n}^{2}$ gives the success probability of the protocol. In other words, the normalization constant is related to the success probability by $\mathcal{N}=1/\sqrt{P}$.
Application of $k$ PR operations is mathematically equivalent to $k$ times multiplication of the effective operator $(\mathbbm{1}\otimes\hat{R}^k)\ket{\Psi_{\textsc{tmsv}}}$. It can be proved by mathematical induction that since PR conserves the photon number of the input mode, the results do not depend on the arrangement of the operations.
That is, so long as TMSV states are concerned, performing all of the PR operations on one of the modes is equivalent to applying them equally on both of the modes:
\begin{equation}
\ket{\Psi_k}\equiv (\mathbbm{1}\otimes\hat{R}^k)\ket{\Psi_{\textsc{tmsv}}}
=(\hat{R}^l\otimes\hat{R}^{k-l})\ket{\Psi_{\textsc{tmsv}}}.
\end{equation}
After performing $k$ PR operations on the TMSV state in Eq.~(\ref{TMSV}) We arrive at
\begin{align}
\ket{\Psi_{k}}&=\mathcal{N}_{k}\sqrt{1-\lambda^{2}}\sum_{n=0}^{\infty} \lambda^{n} T^{k(n-1)} \nonumber\\
&~~~~~\times\left[T^{2}-n(1-T^{2})\right]^{k} \ket{n,n},
\label{CPR}
\end{align} 
where $\mathcal{N}_{k}$ is the corresponding normalization factor.
In calculating the above equation we have assumed that all beam-splitters have the same transmissivity $T$.
The success probability of a CPR protocol consisting of $k$ operations reduces to a closed form series
\begin{equation}
	P_{k}=\sum_{m=0}^{2k} \binom{2k}{m} T^{2k-2m}(-1)^{m}(1-\lambda^{2})(1-T^2)^{m} a_m,
\label{Success Probability}
\end{equation}
where $a_m$ is given by the following recursion relation
\begin{equation}
	a_{m+1}=\frac{T}{2k}\frac{\partial a_{m}}{\partial T},
\label{a_m}
\end{equation}
with $a_0=(1-{\lambda}^{2} {T}^{2k})^{-1}$.
Therefore, one arrives at
\begin{equation}
a_{m}=a_0^{m+1}\sum_{i, j=0}^{m} i ({\lambda}^{2} {T}^{2k})^{i} (1-{\lambda}^{2} {T}^{2k})^{j}.
\label{a_m}
\end{equation}
It is straightforward to show that $P_k$ is a decreasing function of $k$, i.e., $P_{k+1} \leq P_{k}$ [see Appendix~\ref{appendix1}].
Hence, the amount of success probability descends by increasing the number of replacement operations.

In the reminder of this section we derive analytical expressions for the quantum properties of the outcome states Eq.~(\ref{CPR}).

%===============SubSection===============
\subsection{Entanglement \label{subsec:ent}}
We use logarithmic-negativity~\cite{Vidal2002, Plenio2005} to quantify the degree of entanglement. For a quantum state with density matrix $\rho$ the logarithmic-negativity $E_{N}$ is defined as
\begin{equation}
	E_{N}=\log_{2}(\|\rho^{\Gamma}\|_{1})
\end{equation}
where $\rho^{\Gamma}$ denotes the partial transposition of $\rho$, while $\|\bm{\cdot}\|_{1}$ stands for the trace norm.
Logarithmic negativity of a pure state of the form of (\ref{general_form}) is easily evaluated by
\begin{equation}
E_{N}=2\log_{2}\!\Big(\!\mathcal{N}\sum_{n} \abs{c_{n}}\Big).
\label{log-neg}
\end{equation} 
For a TMSV state calculation of the logarithmic-negativity is straightforward as the series turns into a geometric sum. One, therefore, arrives at
\begin{equation}
E_{N}=\log_{2}\!\Big(\frac{1+\lambda}{1-\lambda} \Big).
\end{equation}
For a $k$-step cascaded protocol, we derive the following formula for the logarithmic-negativity
\begin{equation}
E_{N}\!=\log_{2}\!\Big\{\frac{\big[\sum_{l=0}^{k} \binom{k}{l} T^{k-2l}(-1)^{l}(1-T^2)^{l}b_{l}\big]^{2}}{\sum_{l=0}^{2k} \binom{2k}{l} T^{2k-2l}(-1)^{l}(1-T^2)^{l}a_{l}}\Big\},
\label{log-neg_CPR}
\end{equation}
where  $b_0=(1-{\lambda}{T}^{k})^{-1}$ and $b_{l+1}=\frac{T}{k}\frac{\partial b_{l}}{\partial T}$ which give
\begin{equation}
b_{l}=b_0^{l+1}\sum_{i, j=0}^{l} i ({\lambda} {T}^{k})^{i} (1-{\lambda} {T}^{k})^{j}.
\label{b_m}
\end{equation}
In the next section we show that the maximum entanglement monotonically increases by increasing the number of repetitions $k$. In other words, applying a larger number of replacement operations increases the amount of entanglement in the final state, provided an optimal value for $T$ is chosen.
We also see that the maximum entanglement asymptotically reaches to a saturation value.

%===============SubSection===============
\subsection{Non-Gaussianity \label{subsec:non_G}}
Another quantum property of the output states of a CPR protocol is the degree of their non-Gaussianity. This property becomes important for universal quantum computation with CV systems~\cite{Menicucci2006, Alexander2018, Eshaqi2019}.
There are several methods for measuring non-Gaussianity of a state. Here, we use the relative entropy for the convenience~\cite{Genoni2008, Genoni2010, Marian2013}.
For a given quantum state $\rho$, the non-Gaussianity $\mathcal{G}(\rho)$ is quantified by comparing its entropy to the nearest Gaussian state $\rho_{G}$ whose first and second moments in the system operators are the same as $\rho$. Mathematically, it is
\begin{equation}
\mathcal{G}(\rho)=S[\rho_{G}]-S[\rho],
\label{NonG_General}
\end{equation} 
where $S[\rho]=\Tr{\rho\log_{2}\!\rho}$ denotes the von Neumann entropy. Since the states we are working with are pure, entropy of the original states are vanishing and the non-Gaussianity equals to the von Neumann entropy of the closest Gaussian state $\rho_{G}$.
For a Gaussian state, in turn, the entropy is given by~\cite{Weedbrook2012}
\begin{equation}
\mathcal{G}=g(\nu_{+})+g(\nu_{-}),
\label{NonG_formula}
\end{equation}
where $\nu_{\pm}$ are the symplectic eigenvalues of the covariance matrix of the state $\rho_{G}$ and we have introduced
\begin{equation}
g(z)=\frac{z+1}{2}\log_{2}\!\Big(\frac{z+1}{2}\Big)-\frac{z-1}{2}\log_{2}\!\Big(\frac{z-1}{2}\Big).
\label{g_function}
\end{equation}
In order to evaluate the non-Gaussianity, we first obtain the covariance matrix of the Gaussian state $\rho_G$. A two-mode state has four quadrature operators $\boldsymbol{\hat{r}}=(\hat{x}_{1}, \hat{p}_{1}, \hat{x}_{2},\hat{p}_{2})$ which in terms of the annihilation and creation operators are defined as $\hat{x}_{i}=\hat{a}_{i}+\hat{a}^{\dagger}_{i}$ and $\hat{p}_{i}=i(\hat{a}^{\dagger}_{i}-\hat{a}_{i})$.
The first moment vector is simply attained by evaluating the expectation values of the quadratures $\expval{\bm{\hat{r}}}$.
Meanwhile, the elements of the covariance matrix are the second statistical moments given by
\begin{equation}
V_{ij}=\frac{1}{2}\expval{\Delta\hat{r}_{j}\Delta\hat{r}_{k}+\Delta\hat{r}_{k}\Delta\hat{r}_{j}},
\end{equation} 
where $\Delta\hat{r}_{j}=\hat{r}_{j}-\expval{\hat{r}_{j}}$.
The states resulting from our protocol are of the form of equation (\ref{general_form}). Such states have zero first moments $\expval{\hat{r}_{j}}=0$ and their covariance matrix is $4\times4$ in the following form for a $k$ times CPR
\begin{equation}
 	V_{k}=\left(
 	\begin{array}{c|c}
 	\alpha_{k} \mathbb{I}   & \gamma_{k} \sigma\\
 	\hline
 	\gamma_{k} \sigma       & \alpha_{k} \mathbb{I} 
 	\end{array}
 	\right),
\end{equation}
where we have introduced the $2\times2$ matrices $\mathbb{I}=\textrm{diag}(1,1)$ and $\sigma=\textrm{diag}(1,-1)$.
Moreover, the parameters $\alpha_{k}$ and $\gamma_{k}$ are given by
\begin{subequations}
\begin{align}
\alpha_{k} &= 1+2\sum_{n=0}^{\infty}n c_{n}^{2}, \\
\gamma_{k} &= 2\sum_{n=0}^{\infty}(n+1) c_{n}c_{n+1}.
\end{align}	
\end{subequations}
The symplectic eigenvalues $\nu_{\pm}$ are thus found as~\cite{Weedbrook2012}
\begin{equation}
\nu_{+}=\nu_{-}=\sqrt{\alpha_{k}^{2}-\gamma_{k}^{2}}.
\end{equation}
By substituting the eigenvalues $\nu_{\pm}$ in Eq.~(\ref{NonG_formula}) the amount of non-Gaussianity can be computed.

%≡≡≡≡≡≡≡≡≡≡≡≡≡≡≡SECTION≡≡≡≡≡≡≡≡≡≡≡≡≡≡≡
\section{Results \label{sec:result}}
In this section, we study the performance of a cascaded photon replacement protocol by investigating the enhancement resulted in the output state as well as its success probability and non-Gaussianity.
We explore the protocol properties at various parameter values and compare it to the cascaded photon addition and subtraction protocols.

%~~~~~~~~~~~~~~~FIGURE~~~~~~~~~~~~~~~
\begin{figure}[tb]
\includegraphics[width=\columnwidth]{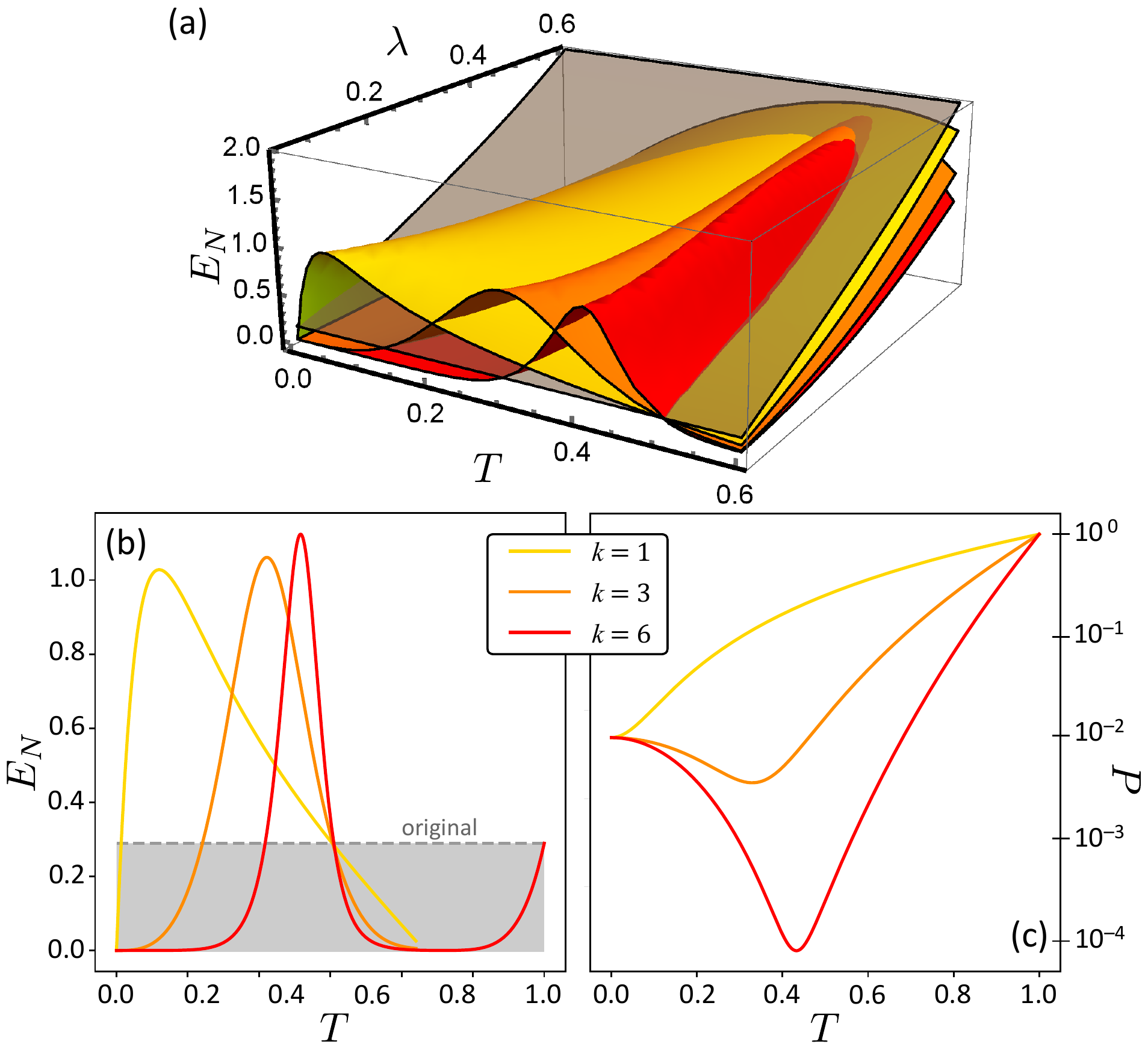}
\caption{\label{CPR_Comparison}%
(a) The logarithmic-negativity of a TMSV (grey transparent surface) as well as the states resulting from applying CPR protocols consisted of 1 (yellow), 3 (orange), and 6 (red) replacement operations.
(b) The plot of logarithmic-negativity against $T$ for $\lambda=0.1$.
(c) The success probability of applying 1, 3, and 6 PR operations for different values of $T$.
The same colors are used in presentation of all diagrams.
}
\end{figure}
%~~~~~~~~~~~~~~~~~~~~~~~~~~~~~~~~~~~~
%~~~~~~~~~~~~~~~FIGURE~~~~~~~~~~~~~~~
\begin{figure}[tb]
\includegraphics[width=0.9\columnwidth]{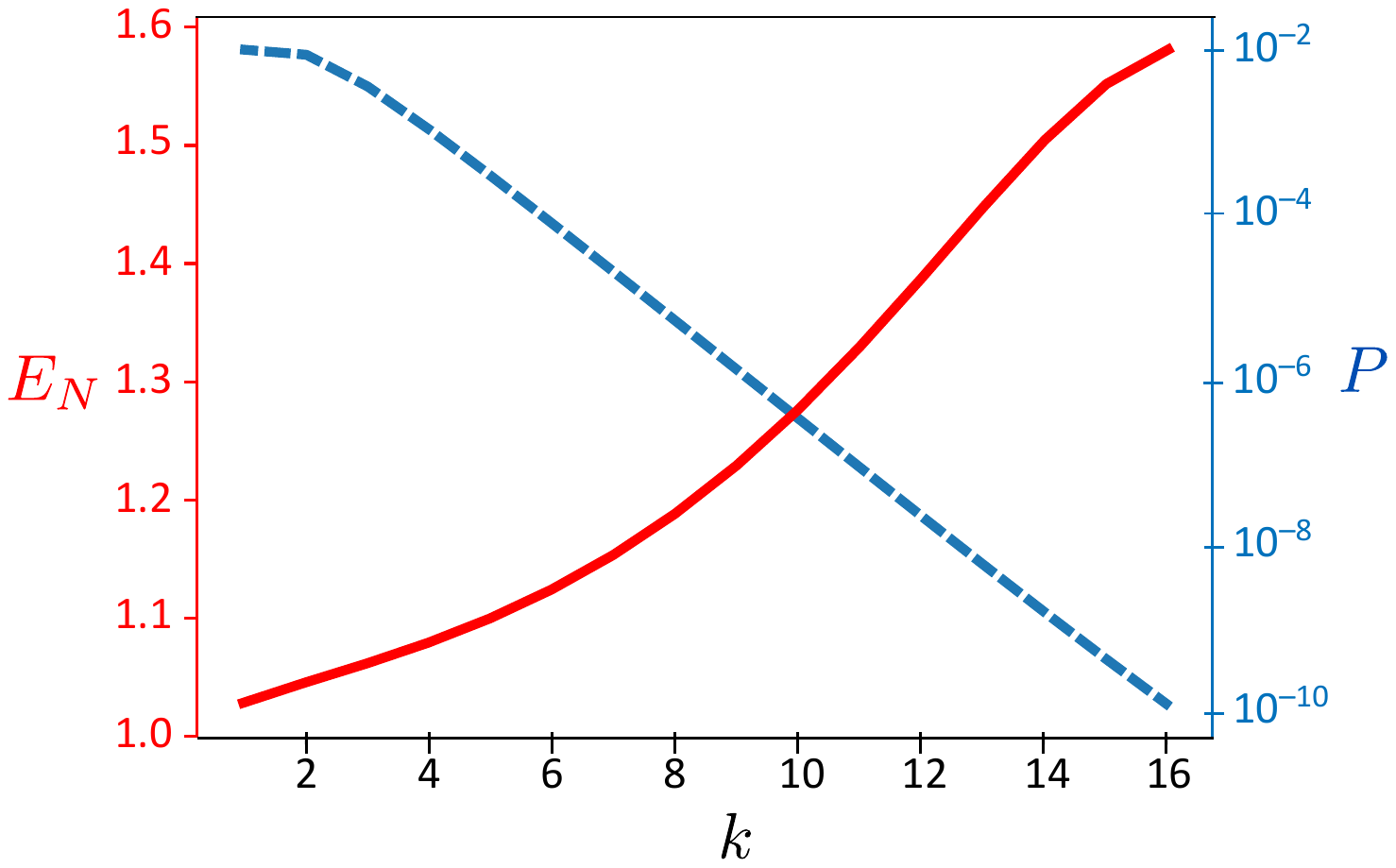}
\caption{\label{trends}%
Trend of the entanglement in a CPR: The maximum distillable entanglement (solid red) and its success probability (dashed blue) as a function of number of repetitions.
}
\end{figure}
%~~~~~~~~~~~~~~~~~~~~~~~~~~~~~~~~~~~~
%~~~~~~~~~~~~~~~FIGURE~~~~~~~~~~~~~~~
\begin{figure*}[t]
\includegraphics[width=.9\textwidth]{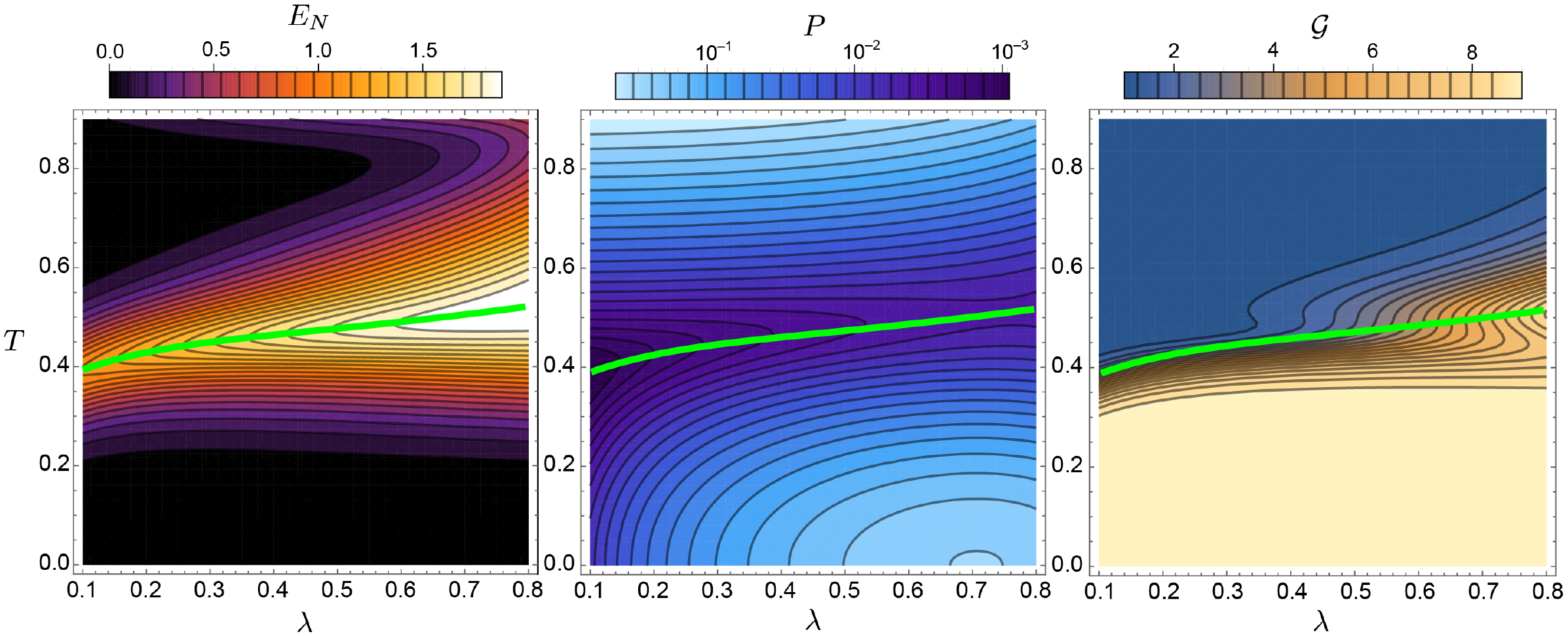}
\caption{\label{PR4_NonG}%
Contour plots of logarithmic-negativity, success probability, and non-Gaussianity of a state distilled by 4 photon replacement operations against the transmissivity of the beam-splitters $T$ and the parameter $\lambda$ that represents the initial squeezing of the TMSV. The green line illustrates locus of the points with maximum entanglement.
}
\end{figure*}
%~~~~~~~~~~~~~~~~~~~~~~~~~~~~~~~~~~~~

%===============SubSection===============
\subsection{Entanglement enhancement}
The main goal of every entanglement distillation protocol is to increase the entanglement of the state.
Therefore, we first examine the degree of enhancement achievable after applying our CPR protocol.
Fig.~\ref{CPR_Comparison}(a) shows logarithmic-negativity of the output state after 1, 3 and 6 CPR operations as a function of initial entanglement $\lambda$ and the filtering transmissivity $T$.
In the figure, the original entanglement of the initial TMSV state is also shown as a reference.
It can be seen that the maximum distillable entanglement is higher than the original value only for weakly entangled initial states ($\lambda \lesssim 0.6$) and in an optimial range of $T$. The optical range and the value of transmissivity that maximizes the entanglement depends on the number of protocol repetitions $k$, so we call it $T_{k,{\rm max}}$.
By increasing the number of operations the amount of maximum entanglement increases. However, the range of transmissivity values in which the distillable entanglement overtakes that of original state gets narrower.
One also notices that the maximum reachable entanglement increases for higher number of repetitions approaches to a 50:50 beam--splitter for very large number of operations.
%$\lim\limits_{k\to\infty}T_{k,{\rm max}} = 0.5$.
At the same time, the bandwidth decreases to zero.
Fig.~\ref{CPR_Comparison}(b) illustrates a cross section of the surfaces at $\lambda=0.1$.
%For odd numbers of photon replacement, some of $c_{n}$ are negative. To simplify the calculations, we restrict values of T to $0 < T < 1/\sqrt{2}$.
%As can be seen in the figure, applying more numbers of PR operations can result in more entanglement if we choose an appropriate $T$.
For $T=0$, where the incident mode does not pass through the beam splitter the entanglement is always vanishing regardless of the number of operations.
On the other hand at $T=1$ the incident single-photon cannot pass the beam-splitter, and thus, does not mix up with the mode. Therefore, the entanglement always equals the original value, as one would expect.
Remarkably, for a 50:50 beam-splitter ($T=0.5$) the entanglement values are independent from the number of replacement operations and equal to that of the original state.
%The reason of this behavior can be traced back to the nature of 
Fig.~\ref{CPR_Comparison}(c) shows the success probability of a state resulting from 1, 3, and 6 PR operations. For a cascaded PR the maximum of the entanglement coincides with the success probability minimum whose value falls exponentially with the number of operations.
%In Fig.~\ref{PR6_Prob} we plot the success probability of the CPR protocol.
%~~~~~~~~~~~~~~~FIGURE~~~~~~~~~~~~~~~
%\begin{figure}[tb]
%\includegraphics[width=0.7\columnwidth]{Figs/P6r}
%\includegraphics[width=0.07\columnwidth]{Figs/P6rcb}
%\caption{\label{PR6_Prob}%
%(a) The contour plot of success probability of the state resulting from applying 6 replacement operations on a TMSV. The $x$-axis is the values of the parameter $\lambda$ which represents the initial squeezing of the TMSV. The $y$-axis is the transmissivity of the beam-splitters $T$. The amount of success probability on each of the contours is written in the boxes. \mac{I would put the $z$-axis in log-scale and use nicer contour labels.}
%\mat{(b)} \mac{The aspect-ratio needs to be adjusted to match the other plot. Also colors similar to that of Fig.~\ref{CPR_Comparison}(b) are recommended.}
%}
%\end{figure}
%~~~~~~~~~~~~~~~~~~~~~~~~~~~~~~~~~~~~

In order to study the trend of enhancement in the entanglement due to our CPR protocol, in Fig.~\ref{trends} we plot logarithmic negativity of the CPR output state against the number of replacement operations.
As anticipate in Sec.~\ref{sec:cpr}A it clearly shows that the amount of entanglement increases by the number of operations.
However, the enhancement is attained with the cost of lower success probabilities as the dashed blue line suggests in Fig.~\ref{trends}.
The entanglement monotonically increases and asymptotically approaches a saturating value.
Meanwhile, the probability that determines the entanglement rate drops down as $P_k\propto 10^{-\frac{2}{3}k}$.
As it will become clear later in this section, this is still a higher value compared to the cascaded photon addition and subtraction success probabilities.

%===============SubSection===============
\subsection{Non-Gaussianity}
We next study the detailed behavior of the quantum properties for a fixed number of operations.
Fig.~\ref{PR4_NonG} shows the density plots of entanglement, probability, and non-Gaussianity of a four step CPR.
First, the entanglement is enhanced only around a 50:50 beam-splitter (the green bold line) and as stated before, only efficient for weak initial entanglements.
One notices that the minimum probability follows the same line. However, the success probability increases as the initial entanglement is increased. This indeed indicates that the cost of entanglement enhancement becomes quite affordable and the entanglement rate becomes reasonably high when $\lambda$ approaches the limit of enhancement.

The right most panel in Fig.~\ref{PR4_NonG} illustrates non-Gaussianity of the state resulting from applying a CPR protocol consisted of 4 replacement operations on a TMSV.
It can be inferred from the plot that the non-Gaussianity of the output state experiences a sudden change where the beam-splitter transmissivity maximizes the entanglement, while the non-Gaussianity in other areas is almost constant.
For $T=1$ the amount of non-Gaussianity measure is vanishing. However, as the probability of single-photon mixing gets increased, the state begins to deGaussify until it jumps to a plateau.
This property can be benefited in the quantum protocols where the amount of non-Gaussianity needs fine tuning.
In particular, the linewidth of the entanglement with respect to $T$ is larger than the slope of $\mathcal{G}$ at small-entanglement regimes. Therefore, one switches from an almost Gaussian state to a highly non-Gaussian one without too much change in the entanglement.

%===============SubSection===============
\subsection{Comparison with cascaded photon addition and photon subtraction}
\begin{figure}[b]
	\includegraphics[width=\columnwidth]{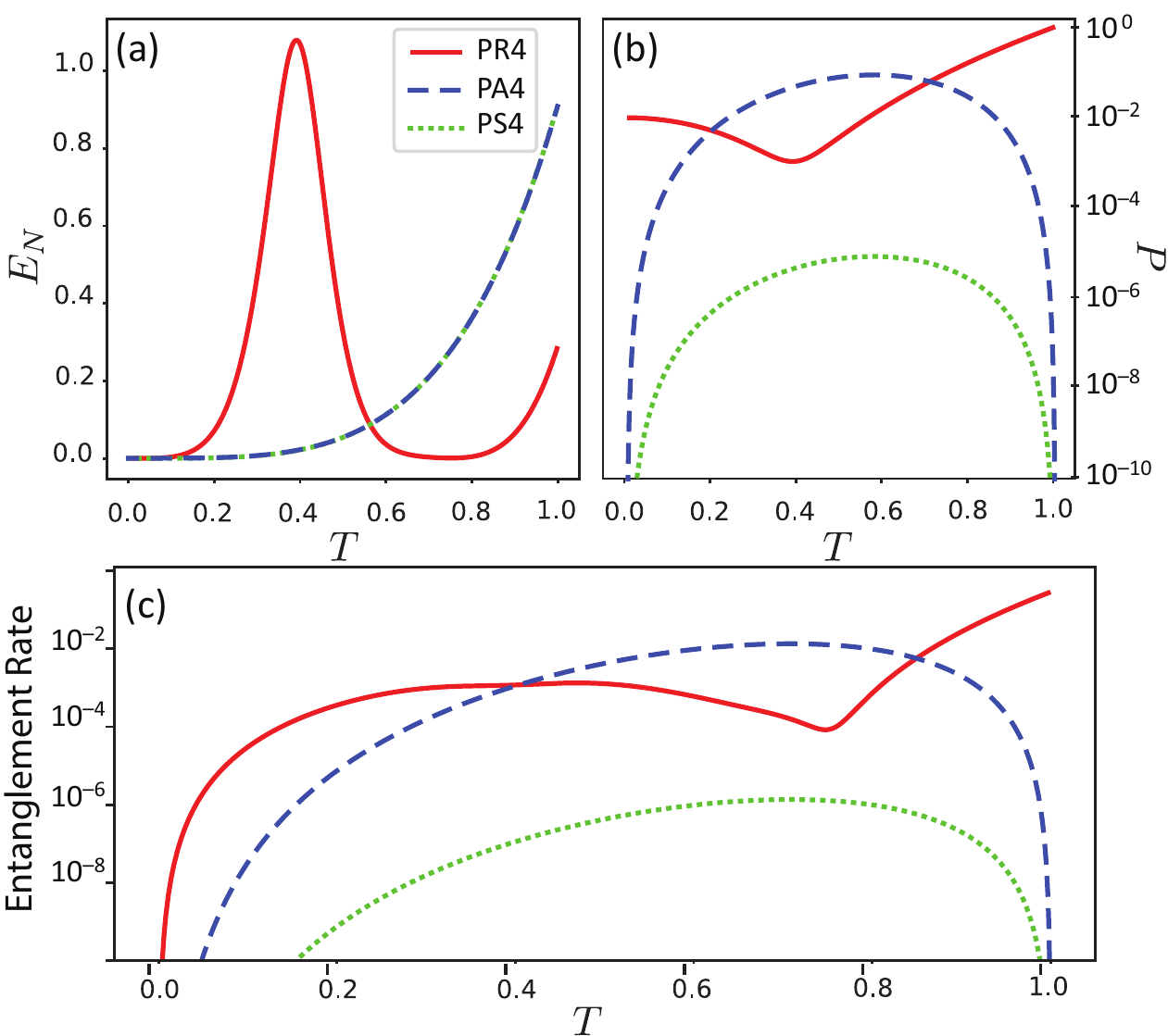}
\caption{\label{Add_Sub_fig}%
Logarithmic-negativity (a), success probability (b), and entanglement rate (c) of the cascaded entanglement distillation protocols: photon replacement (solid line), photon addition (dashed line), and photon subtraction (dotted line).
The results are compared for four-stepped protocol.
}
\end{figure}

We devote this subsection to the comparison of CPR with cascaded photon addition (PA) and subtraction (PS) protocols. The PA and PS cases have been studied for ideal operations in Ref.~\cite{Navarrete2012} and the non-ideal case of single- and two-mode operations in Ref.~\cite{Bartley2015}.
Here, we assume that each operation is performed with finite transmissivity for the beam-splitters.
The results are summarized in Fig.~\ref{Add_Sub_fig}, where we show the entanglement and success probabilities for four-step cascades of PR, PA, and PS.
Since for the addition and subtraction protocols the arrangement of operations is important, here we only consider the symmetric case. That is, two operations on each mode. For this case, the variations of entanglement with $T$ is the same for both cascaded addition and subtraction. The entanglement reaches its maximum value only at the full beam-splitter transmissivity $T=1$.
Note that close this point the success probability is very small and there is a very small probability for the SPD of giving the favoring output (no-click for PA and click for PS). Hence, the entanglement rates (product of the amount of entanglement and the success probability) assume very small values.
The success probability is the lowest where the maximum entanglement is achieved in all cases. Nevertheless, the multiplication of logarithmic negativity and the success probability, which is a measure of the entanglement rate gives a better understanding about the three protocols. Therefore, in Fig.~\ref{Add_Sub_fig}(c) we plot this quantity. We exclude the parameter region around $T\simeq 1$ in the following analysis because the output of a CPR is same as the input. In a wide parameter region that the non-Gaussianity is higher the CPR dominates the other two. Meanwhile the cascaded photon addition provides a better performance result over the rest of region up to the values close to unity, where onwards ($0.9\lesssim T<1$) the CPR overtakes again.

%~~~~~~~~~~~~~~~FIGURE~~~~~~~~~~~~~~~
%~~~~~~~~~~~~~~~~~~~~~~~~~~~~~~~~~~~~

%≡≡≡≡≡≡≡≡≡≡≡≡≡≡≡SECTION≡≡≡≡≡≡≡≡≡≡≡≡≡≡≡
\section{Conclusion \label{sec:concl}}
In summary we have studied the performance of a cascaded photon replacement protocol.
We have shown that a cascade of PR operations can enhance the entanglement of an input TMSV state, provided the initial squeezing is not strong ($\lambda \lesssim 0.6$).
The maximum available entanglement saturates to an asymptotical value as the number operations increases. The success probability, however, decreases exponentially.
The non-Gaussian properties of the resulting states exhibit sensitivity to the beam-splitter transmissivity $T$, which is significant for weak initial entanglements.
The value of non-Gaussianity of the output state drops down from a plateau to almost a Gaussian state over a short range around $T_{\rm max}$, while the entanglement retains its value.
Our studies show that CPR for pure states is insensitive to the arrangement of the PR operations on the modes.
This asymmetric nature of our protocol allows for enhancing the entanglement by only operating on one of the modes. This can prove beneficial for setups where all or parts of the PR operation are not experimentally feasible for one of the modes while it is available for the other mode. 
For example, one enhances the microwave-optical entanglement in a hybrid system~\cite{Barzanjeh2012, Andrews2014} or the optomechanical entanglement~\cite{Palomaki2013, Vitali2007, Genes2008, Abdi2012a} only by operating on the optical parties.
\bibliography{CasPR}

%apsrev4-2.bst 2019-01-14 (MD) hand-edited version of apsrev4-1.bst
%Control: key (0)
%Control: author (8) initials jnrlst
%Control: editor formatted (1) identically to author
%Control: production of article title (0) allowed
%Control: page (0) single
%Control: year (1) truncated
%Control: production of eprint (0) enabled
\begin{thebibliography}{47}%
\makeatletter
\providecommand \@ifxundefined [1]{%
 \@ifx{#1\undefined}
}%
\providecommand \@ifnum [1]{%
 \ifnum #1\expandafter \@firstoftwo
 \else \expandafter \@secondoftwo
 \fi
}%
\providecommand \@ifx [1]{%
 \ifx #1\expandafter \@firstoftwo
 \else \expandafter \@secondoftwo
 \fi
}%
\providecommand \natexlab [1]{#1}%
\providecommand \enquote  [1]{``#1''}%
\providecommand \bibnamefont  [1]{#1}%
\providecommand \bibfnamefont [1]{#1}%
\providecommand \citenamefont [1]{#1}%
\providecommand \href@noop [0]{\@secondoftwo}%
\providecommand \href [0]{\begingroup \@sanitize@url \@href}%
\providecommand \@href[1]{\@@startlink{#1}\@@href}%
\providecommand \@@href[1]{\endgroup#1\@@endlink}%
\providecommand \@sanitize@url [0]{\catcode `\\12\catcode `\$12\catcode
  `\&12\catcode `\#12\catcode `\^12\catcode `\_12\catcode `\%12\relax}%
\providecommand \@@startlink[1]{}%
\providecommand \@@endlink[0]{}%
\providecommand \url  [0]{\begingroup\@sanitize@url \@url }%
\providecommand \@url [1]{\endgroup\@href {#1}{\urlprefix }}%
\providecommand \urlprefix  [0]{URL }%
\providecommand \Eprint [0]{\href }%
\providecommand \doibase [0]{https://doi.org/}%
\providecommand \selectlanguage [0]{\@gobble}%
\providecommand \bibinfo  [0]{\@secondoftwo}%
\providecommand \bibfield  [0]{\@secondoftwo}%
\providecommand \translation [1]{[#1]}%
\providecommand \BibitemOpen [0]{}%
\providecommand \bibitemStop [0]{}%
\providecommand \bibitemNoStop [0]{.\EOS\space}%
\providecommand \EOS [0]{\spacefactor3000\relax}%
\providecommand \BibitemShut  [1]{\csname bibitem#1\endcsname}%
\let\auto@bib@innerbib\@empty
%</preamble>
\bibitem [{\citenamefont {Steane}(1999)}]{Steane1999}%
  \BibitemOpen
  \bibfield  {author} {\bibinfo {author} {\bibfnamefont {A.~M.}\ \bibnamefont
  {Steane}},\ }\bibfield  {title} {\bibinfo {title} {Efficient fault-tolerant
  quantum computing},\ }\href {https://doi.org/10.1038/20127} {\bibfield
  {journal} {\bibinfo  {journal} {Nature}\ }\textbf {\bibinfo {volume} {399}},\
  \bibinfo {pages} {124} (\bibinfo {year} {1999})}\BibitemShut {NoStop}%
\bibitem [{\citenamefont {Bennett}\ \emph
  {et~al.}(1996{\natexlab{a}})\citenamefont {Bennett}, \citenamefont
  {Brassard}, \citenamefont {Popescu}, \citenamefont {Schumacher},
  \citenamefont {Smolin},\ and\ \citenamefont {Wootters}}]{Bennett1996}%
  \BibitemOpen
  \bibfield  {author} {\bibinfo {author} {\bibfnamefont {C.~H.}\ \bibnamefont
  {Bennett}}, \bibinfo {author} {\bibfnamefont {G.}~\bibnamefont {Brassard}},
  \bibinfo {author} {\bibfnamefont {S.}~\bibnamefont {Popescu}}, \bibinfo
  {author} {\bibfnamefont {B.}~\bibnamefont {Schumacher}}, \bibinfo {author}
  {\bibfnamefont {J.~A.}\ \bibnamefont {Smolin}},\ and\ \bibinfo {author}
  {\bibfnamefont {W.~K.}\ \bibnamefont {Wootters}},\ }\bibfield  {title}
  {\bibinfo {title} {Purification of noisy entanglement and faithful
  teleportation via noisy channels},\ }\href
  {https://doi.org/10.1103/PhysRevLett.76.722} {\bibfield  {journal} {\bibinfo
  {journal} {Phys. Rev. Lett.}\ }\textbf {\bibinfo {volume} {76}},\ \bibinfo
  {pages} {722} (\bibinfo {year} {1996}{\natexlab{a}})}\BibitemShut {NoStop}%
\bibitem [{\citenamefont {Bennett}\ \emph
  {et~al.}(1996{\natexlab{b}})\citenamefont {Bennett}, \citenamefont
  {Bernstein}, \citenamefont {Popescu},\ and\ \citenamefont
  {Schumacher}}]{Bennett1996a}%
  \BibitemOpen
  \bibfield  {author} {\bibinfo {author} {\bibfnamefont {C.~H.}\ \bibnamefont
  {Bennett}}, \bibinfo {author} {\bibfnamefont {H.~J.}\ \bibnamefont
  {Bernstein}}, \bibinfo {author} {\bibfnamefont {S.}~\bibnamefont {Popescu}},\
  and\ \bibinfo {author} {\bibfnamefont {B.}~\bibnamefont {Schumacher}},\
  }\bibfield  {title} {\bibinfo {title} {{Concentrating partial entanglement by
  local operations}},\ }\href {https://doi.org/10.1103/PhysRevA.53.2046}
  {\bibfield  {journal} {\bibinfo  {journal} {Phys. Rev. A}\ }\textbf {\bibinfo
  {volume} {53}},\ \bibinfo {pages} {2046} (\bibinfo {year}
  {1996}{\natexlab{b}})}\BibitemShut {NoStop}%
\bibitem [{\citenamefont {Giedke}\ \emph {et~al.}(2001)\citenamefont {Giedke},
  \citenamefont {Kraus}, \citenamefont {Lewenstein},\ and\ \citenamefont
  {Cirac}}]{Giedke2001}%
  \BibitemOpen
  \bibfield  {author} {\bibinfo {author} {\bibfnamefont {G.}~\bibnamefont
  {Giedke}}, \bibinfo {author} {\bibfnamefont {B.}~\bibnamefont {Kraus}},
  \bibinfo {author} {\bibfnamefont {M.}~\bibnamefont {Lewenstein}},\ and\
  \bibinfo {author} {\bibfnamefont {J.~I.}\ \bibnamefont {Cirac}},\ }\bibfield
  {title} {\bibinfo {title} {Entanglement criteria for all bipartite gaussian
  states},\ }\href {https://doi.org/10.1103/PhysRevLett.87.167904} {\bibfield
  {journal} {\bibinfo  {journal} {Phys. Rev. Lett.}\ }\textbf {\bibinfo
  {volume} {87}},\ \bibinfo {pages} {167904} (\bibinfo {year}
  {2001})}\BibitemShut {NoStop}%
\bibitem [{\citenamefont {Giedke}\ and\ \citenamefont {{Ignacio
  Cirac}}(2002)}]{Giedke2002}%
  \BibitemOpen
  \bibfield  {author} {\bibinfo {author} {\bibfnamefont {G.}~\bibnamefont
  {Giedke}}\ and\ \bibinfo {author} {\bibfnamefont {J.}~\bibnamefont {{Ignacio
  Cirac}}},\ }\bibfield  {title} {\bibinfo {title} {{Characterization of
  Gaussian operations and distillation of Gaussian states}},\ }\href
  {https://doi.org/10.1103/PhysRevA.66.032316} {\bibfield  {journal} {\bibinfo
  {journal} {Phys. Rev. A}\ }\textbf {\bibinfo {volume} {66}},\ \bibinfo
  {pages} {032316} (\bibinfo {year} {2002})}\BibitemShut {NoStop}%
\bibitem [{\citenamefont {Fiur{\'{a}}{\v{s}}ek}(2002)}]{Fiurasek2002}%
  \BibitemOpen
  \bibfield  {author} {\bibinfo {author} {\bibfnamefont {J.}~\bibnamefont
  {Fiur{\'{a}}{\v{s}}ek}},\ }\bibfield  {title} {\bibinfo {title} {{Gaussian
  Transformations and Distillation of Entangled Gaussian States}},\ }\href
  {https://doi.org/10.1103/PhysRevLett.89.137904} {\bibfield  {journal}
  {\bibinfo  {journal} {Phys. Rev. Lett.}\ }\textbf {\bibinfo {volume} {89}},\
  \bibinfo {pages} {137904} (\bibinfo {year} {2002})}\BibitemShut {NoStop}%
\bibitem [{\citenamefont {Eisert}\ \emph {et~al.}(2002)\citenamefont {Eisert},
  \citenamefont {Scheel},\ and\ \citenamefont {Plenio}}]{Eisert2002}%
  \BibitemOpen
  \bibfield  {author} {\bibinfo {author} {\bibfnamefont {J.}~\bibnamefont
  {Eisert}}, \bibinfo {author} {\bibfnamefont {S.}~\bibnamefont {Scheel}},\
  and\ \bibinfo {author} {\bibfnamefont {M.~B.}\ \bibnamefont {Plenio}},\
  }\bibfield  {title} {\bibinfo {title} {{Distilling Gaussian States with
  Gaussian Operations is Impossible}},\ }\href
  {https://doi.org/10.1103/PhysRevLett.89.137903} {\bibfield  {journal}
  {\bibinfo  {journal} {Phys. Rev. Lett.}\ }\textbf {\bibinfo {volume} {89}},\
  \bibinfo {pages} {137903} (\bibinfo {year} {2002})}\BibitemShut {NoStop}%
\bibitem [{\citenamefont {Opatrn{\'{y}}}\ \emph {et~al.}(2000)\citenamefont
  {Opatrn{\'{y}}}, \citenamefont {Kurizki},\ and\ \citenamefont
  {Welsch}}]{Opatrny2000}%
  \BibitemOpen
  \bibfield  {author} {\bibinfo {author} {\bibfnamefont {T.}~\bibnamefont
  {Opatrn{\'{y}}}}, \bibinfo {author} {\bibfnamefont {G.}~\bibnamefont
  {Kurizki}},\ and\ \bibinfo {author} {\bibfnamefont {D.~G.}\ \bibnamefont
  {Welsch}},\ }\bibfield  {title} {\bibinfo {title} {{Improvement on
  teleportation of continuous variables by photon subtraction via conditional
  measurement}},\ }\href {https://doi.org/10.1103/PhysRevA.61.032302}
  {\bibfield  {journal} {\bibinfo  {journal} {Phys. Rev. A}\ }\textbf {\bibinfo
  {volume} {61}},\ \bibinfo {pages} {032302} (\bibinfo {year}
  {2000})}\BibitemShut {NoStop}%
\bibitem [{\citenamefont {Kitagawa}\ \emph {et~al.}(2006)\citenamefont
  {Kitagawa}, \citenamefont {Takeoka}, \citenamefont {Sasaki},\ and\
  \citenamefont {Chefles}}]{Kitagawa2006}%
  \BibitemOpen
  \bibfield  {author} {\bibinfo {author} {\bibfnamefont {A.}~\bibnamefont
  {Kitagawa}}, \bibinfo {author} {\bibfnamefont {M.}~\bibnamefont {Takeoka}},
  \bibinfo {author} {\bibfnamefont {M.}~\bibnamefont {Sasaki}},\ and\ \bibinfo
  {author} {\bibfnamefont {A.}~\bibnamefont {Chefles}},\ }\bibfield  {title}
  {\bibinfo {title} {{Entanglement evaluation of non-Gaussian states generated
  by photon subtraction from squeezed states}},\ }\href
  {https://doi.org/10.1103/PhysRevA.73.042310} {\bibfield  {journal} {\bibinfo
  {journal} {Phys. Rev. A}\ }\textbf {\bibinfo {volume} {73}},\ \bibinfo
  {pages} {042310} (\bibinfo {year} {2006})}\BibitemShut {NoStop}%
\bibitem [{\citenamefont {Ourjoumtsev}\ \emph {et~al.}(2007)\citenamefont
  {Ourjoumtsev}, \citenamefont {Dantan}, \citenamefont {Tualle-Brouri},\ and\
  \citenamefont {Grangier}}]{Ourjoumtsev2007}%
  \BibitemOpen
  \bibfield  {author} {\bibinfo {author} {\bibfnamefont {A.}~\bibnamefont
  {Ourjoumtsev}}, \bibinfo {author} {\bibfnamefont {A.}~\bibnamefont {Dantan}},
  \bibinfo {author} {\bibfnamefont {R.}~\bibnamefont {Tualle-Brouri}},\ and\
  \bibinfo {author} {\bibfnamefont {P.}~\bibnamefont {Grangier}},\ }\bibfield
  {title} {\bibinfo {title} {{Increasing Entanglement between Gaussian States
  by Coherent Photon Subtraction}},\ }\href
  {https://doi.org/10.1103/PhysRevLett.98.030502} {\bibfield  {journal}
  {\bibinfo  {journal} {Phys. Rev. Lett.}\ }\textbf {\bibinfo {volume} {98}},\
  \bibinfo {pages} {030502} (\bibinfo {year} {2007})}\BibitemShut {NoStop}%
\bibitem [{\citenamefont {Takahashi}\ \emph {et~al.}(2010)\citenamefont
  {Takahashi}, \citenamefont {Neergaard-nielsen}, \citenamefont {Takeuchi},
  \citenamefont {Takeoka}, \citenamefont {Hayasaka}, \citenamefont {Furusawa},\
  and\ \citenamefont {Sasaki}}]{Takahashi2010}%
  \BibitemOpen
  \bibfield  {author} {\bibinfo {author} {\bibfnamefont {H.}~\bibnamefont
  {Takahashi}}, \bibinfo {author} {\bibfnamefont {J.~S.}\ \bibnamefont
  {Neergaard-nielsen}}, \bibinfo {author} {\bibfnamefont {M.}~\bibnamefont
  {Takeuchi}}, \bibinfo {author} {\bibfnamefont {M.}~\bibnamefont {Takeoka}},
  \bibinfo {author} {\bibfnamefont {K.}~\bibnamefont {Hayasaka}}, \bibinfo
  {author} {\bibfnamefont {A.}~\bibnamefont {Furusawa}},\ and\ \bibinfo
  {author} {\bibfnamefont {M.}~\bibnamefont {Sasaki}},\ }\bibfield  {title}
  {\bibinfo {title} {{Entanglement distillation from Gaussian input states}},\
  }\href {https://doi.org/10.1038/nphoton.2010.1} {\bibfield  {journal}
  {\bibinfo  {journal} {Nat. Photon.}\ }\textbf {\bibinfo {volume} {4}},\
  \bibinfo {pages} {178} (\bibinfo {year} {2010})}\BibitemShut {NoStop}%
\bibitem [{\citenamefont {Zhang}\ and\ \citenamefont {van
  Loock}(2010)}]{Zhang2010}%
  \BibitemOpen
  \bibfield  {author} {\bibinfo {author} {\bibfnamefont {S.~L.}\ \bibnamefont
  {Zhang}}\ and\ \bibinfo {author} {\bibfnamefont {P.}~\bibnamefont {van
  Loock}},\ }\bibfield  {title} {\bibinfo {title} {{Distillation of mixed-state
  continuous-variable entanglement by photon subtraction}},\ }\href
  {https://doi.org/10.1103/PhysRevA.82.062316} {\bibfield  {journal} {\bibinfo
  {journal} {Phys. Rev. A}\ }\textbf {\bibinfo {volume} {82}},\ \bibinfo
  {pages} {062316} (\bibinfo {year} {2010})}\BibitemShut {NoStop}%
\bibitem [{\citenamefont {Zhang}\ \emph {et~al.}(2013)\citenamefont {Zhang},
  \citenamefont {Dong}, \citenamefont {Zou}, \citenamefont {Shi},\ and\
  \citenamefont {Guo}}]{Zhang2013}%
  \BibitemOpen
  \bibfield  {author} {\bibinfo {author} {\bibfnamefont {S.~L.}\ \bibnamefont
  {Zhang}}, \bibinfo {author} {\bibfnamefont {Y.~L.}\ \bibnamefont {Dong}},
  \bibinfo {author} {\bibfnamefont {X.~B.}\ \bibnamefont {Zou}}, \bibinfo
  {author} {\bibfnamefont {B.~S.}\ \bibnamefont {Shi}},\ and\ \bibinfo {author}
  {\bibfnamefont {G.~C.}\ \bibnamefont {Guo}},\ }\bibfield  {title} {\bibinfo
  {title} {{Continuous-variable-entanglement distillation with photon
  addition}},\ }\href {https://doi.org/10.1103/PhysRevA.88.032324} {\bibfield
  {journal} {\bibinfo  {journal} {Phys. Rev. A}\ }\textbf {\bibinfo {volume}
  {88}},\ \bibinfo {pages} {032324} (\bibinfo {year} {2013})}\BibitemShut
  {NoStop}%
\bibitem [{\citenamefont {Lvovsky}\ and\ \citenamefont
  {Mlynek}(2002)}]{Lvovsky2002}%
  \BibitemOpen
  \bibfield  {author} {\bibinfo {author} {\bibfnamefont {A.~I.}\ \bibnamefont
  {Lvovsky}}\ and\ \bibinfo {author} {\bibfnamefont {J.}~\bibnamefont
  {Mlynek}},\ }\bibfield  {title} {\bibinfo {title} {{Quantum-Optical
  Catalysis: Generating Nonclassical States of Light by Means of Linear
  Optics}},\ }\href {https://doi.org/10.1103/PhysRevLett.88.250401} {\bibfield
  {journal} {\bibinfo  {journal} {Phys. Rev. Lett.}\ }\textbf {\bibinfo
  {volume} {88}},\ \bibinfo {pages} {250401} (\bibinfo {year}
  {2002})}\BibitemShut {NoStop}%
\bibitem [{\citenamefont {Lami}\ \emph {et~al.}()\citenamefont {Lami},
  \citenamefont {Takagi},\ and\ \citenamefont {Adesso}}]{Adesso2019}%
  \BibitemOpen
  \bibfield  {author} {\bibinfo {author} {\bibfnamefont {L.}~\bibnamefont
  {Lami}}, \bibinfo {author} {\bibfnamefont {R.}~\bibnamefont {Takagi}},\ and\
  \bibinfo {author} {\bibfnamefont {G.}~\bibnamefont {Adesso}},\ }\bibfield
  {title} {\bibinfo {title} {{Assisted distillation of Gaussian resources}},\
  }\href@noop {} {\ }\Eprint {https://arxiv.org/abs/1905.02173}
  {arXiv:1905.02173 [quant-ph]} \BibitemShut {NoStop}%
\bibitem [{\citenamefont {Yang}\ and\ \citenamefont {Li}(2009)}]{Yang2009}%
  \BibitemOpen
  \bibfield  {author} {\bibinfo {author} {\bibfnamefont {Y.}~\bibnamefont
  {Yang}}\ and\ \bibinfo {author} {\bibfnamefont {F.-L.}\ \bibnamefont {Li}},\
  }\bibfield  {title} {\bibinfo {title} {{Entanglement properties of
  non-Gaussian resources generated via photon subtraction and addition and
  continuous-variable quantum-teleportation improvement}},\ }\href
  {https://doi.org/10.1103/PhysRevA.80.022315} {\bibfield  {journal} {\bibinfo
  {journal} {Phys. Rev. A}\ }\textbf {\bibinfo {volume} {80}},\ \bibinfo
  {pages} {022315} (\bibinfo {year} {2009})}\BibitemShut {NoStop}%
\bibitem [{\citenamefont {Lee}\ \emph {et~al.}(2011)\citenamefont {Lee},
  \citenamefont {Ji}, \citenamefont {Kim},\ and\ \citenamefont
  {Nha}}]{Nha2011}%
  \BibitemOpen
  \bibfield  {author} {\bibinfo {author} {\bibfnamefont {S.~Y.}\ \bibnamefont
  {Lee}}, \bibinfo {author} {\bibfnamefont {S.~W.}\ \bibnamefont {Ji}},
  \bibinfo {author} {\bibfnamefont {H.~J.}\ \bibnamefont {Kim}},\ and\ \bibinfo
  {author} {\bibfnamefont {H.}~\bibnamefont {Nha}},\ }\bibfield  {title}
  {\bibinfo {title} {{Enhancing quantum entanglement for continuous variables
  by a coherent superposition of photon subtraction and addition}},\ }\href
  {https://doi.org/10.1103/PhysRevA.84.012302} {\bibfield  {journal} {\bibinfo
  {journal} {Phys. Rev. A}\ }\textbf {\bibinfo {volume} {84}},\ \bibinfo
  {pages} {012302} (\bibinfo {year} {2011})}\BibitemShut {NoStop}%
\bibitem [{\citenamefont {Wu}\ \emph {et~al.}(2015)\citenamefont {Wu},
  \citenamefont {Liu}, \citenamefont {Hu}, \citenamefont {Huang}, \citenamefont
  {Duan},\ and\ \citenamefont {Ji}}]{Liu2015}%
  \BibitemOpen
  \bibfield  {author} {\bibinfo {author} {\bibfnamefont {J.}~\bibnamefont
  {Wu}}, \bibinfo {author} {\bibfnamefont {S.}~\bibnamefont {Liu}}, \bibinfo
  {author} {\bibfnamefont {L.}~\bibnamefont {Hu}}, \bibinfo {author}
  {\bibfnamefont {J.}~\bibnamefont {Huang}}, \bibinfo {author} {\bibfnamefont
  {Z.}~\bibnamefont {Duan}},\ and\ \bibinfo {author} {\bibfnamefont
  {Y.}~\bibnamefont {Ji}},\ }\bibfield  {title} {\bibinfo {title} {{Improving
  entanglement of even entangled coherent states by a coherent superposition of
  photon subtraction and addition}},\ }\href
  {https://doi.org/10.1364/JOSAB.32.002299} {\bibfield  {journal} {\bibinfo
  {journal} {J. Opt. Soc. Am. B}\ }\textbf {\bibinfo {volume} {32}},\ \bibinfo
  {pages} {2299} (\bibinfo {year} {2015})}\BibitemShut {NoStop}%
\bibitem [{\citenamefont {Navarrete-Benlloch}\ \emph
  {et~al.}(2012)\citenamefont {Navarrete-Benlloch}, \citenamefont
  {Garc{\'{i}}a-Patr{\'{o}}n}, \citenamefont {Shapiro},\ and\ \citenamefont
  {Cerf}}]{Navarrete2012}%
  \BibitemOpen
  \bibfield  {author} {\bibinfo {author} {\bibfnamefont {C.}~\bibnamefont
  {Navarrete-Benlloch}}, \bibinfo {author} {\bibfnamefont {R.}~\bibnamefont
  {Garc{\'{i}}a-Patr{\'{o}}n}}, \bibinfo {author} {\bibfnamefont {J.~H.}\
  \bibnamefont {Shapiro}},\ and\ \bibinfo {author} {\bibfnamefont {N.~J.}\
  \bibnamefont {Cerf}},\ }\bibfield  {title} {\bibinfo {title} {{Enhancing
  quantum entanglement by photon addition and subtraction}},\ }\href
  {https://doi.org/10.1103/PhysRevA.86.012328} {\bibfield  {journal} {\bibinfo
  {journal} {Phys. Rev. A}\ }\textbf {\bibinfo {volume} {86}},\ \bibinfo
  {pages} {012328} (\bibinfo {year} {2012})}\BibitemShut {NoStop}%
\bibitem [{\citenamefont {Bartley}\ and\ \citenamefont
  {Walmsley}(2015)}]{Bartley2015}%
  \BibitemOpen
  \bibfield  {author} {\bibinfo {author} {\bibfnamefont {T.~J.}\ \bibnamefont
  {Bartley}}\ and\ \bibinfo {author} {\bibfnamefont {I.~A.}\ \bibnamefont
  {Walmsley}},\ }\bibfield  {title} {\bibinfo {title} {{Directly comparing
  entanglement-enhancing non-Gaussian operations}},\ }\href
  {https://doi.org/10.1088/1367-2630/17/2/023038} {\bibfield  {journal}
  {\bibinfo  {journal} {New J. Phys.}\ }\textbf {\bibinfo {volume} {17}},\
  \bibinfo {pages} {23038} (\bibinfo {year} {2015})}\BibitemShut {NoStop}%
\bibitem [{\citenamefont {Xu}(2015)}]{Xu2015}%
  \BibitemOpen
  \bibfield  {author} {\bibinfo {author} {\bibfnamefont {X.~X.}\ \bibnamefont
  {Xu}},\ }\bibfield  {title} {\bibinfo {title} {{Enhancing quantum
  entanglement and quantum teleportation for two-mode squeezed vacuum state by
  local quantum-optical catalysis}},\ }\href
  {https://doi.org/10.1103/PhysRevA.92.012318} {\bibfield  {journal} {\bibinfo
  {journal} {Phys. Rev. A}\ }\textbf {\bibinfo {volume} {92}},\ \bibinfo
  {pages} {012318} (\bibinfo {year} {2015})}\BibitemShut {NoStop}%
\bibitem [{\citenamefont {Hosseinidehaj}\ and\ \citenamefont
  {Malaney}(2015)}]{Hosseinidehaj2015}%
  \BibitemOpen
  \bibfield  {author} {\bibinfo {author} {\bibfnamefont {N.}~\bibnamefont
  {Hosseinidehaj}}\ and\ \bibinfo {author} {\bibfnamefont {R.}~\bibnamefont
  {Malaney}},\ }\bibfield  {title} {\bibinfo {title} {Entanglement generation
  via non-gaussian transfer over atmospheric fading channels},\ }\href
  {https://doi.org/10.1103/PhysRevA.92.062336} {\bibfield  {journal} {\bibinfo
  {journal} {Phys. Rev. A}\ }\textbf {\bibinfo {volume} {92}},\ \bibinfo
  {pages} {062336} (\bibinfo {year} {2015})}\BibitemShut {NoStop}%
\bibitem [{\citenamefont {Browne}\ \emph {et~al.}(2003)\citenamefont {Browne},
  \citenamefont {Eisert}, \citenamefont {Scheel},\ and\ \citenamefont
  {Plenio}}]{Browne2003}%
  \BibitemOpen
  \bibfield  {author} {\bibinfo {author} {\bibfnamefont {D.~E.}\ \bibnamefont
  {Browne}}, \bibinfo {author} {\bibfnamefont {J.}~\bibnamefont {Eisert}},
  \bibinfo {author} {\bibfnamefont {S.}~\bibnamefont {Scheel}},\ and\ \bibinfo
  {author} {\bibfnamefont {M.~B.}\ \bibnamefont {Plenio}},\ }\bibfield  {title}
  {\bibinfo {title} {Driving non-gaussian to gaussian states with linear
  optics},\ }\href {https://doi.org/10.1103/physreva.67.062320} {\bibfield
  {journal} {\bibinfo  {journal} {Phys. Rev. A}\ }\textbf {\bibinfo {volume}
  {67}},\ \bibinfo {pages} {062320} (\bibinfo {year} {2003})}\BibitemShut
  {NoStop}%
\bibitem [{\citenamefont {Eisert}\ \emph {et~al.}(2004)\citenamefont {Eisert},
  \citenamefont {Browne}, \citenamefont {Scheel},\ and\ \citenamefont
  {Plenio}}]{Eisert2004}%
  \BibitemOpen
  \bibfield  {author} {\bibinfo {author} {\bibfnamefont {J.}~\bibnamefont
  {Eisert}}, \bibinfo {author} {\bibfnamefont {D.}~\bibnamefont {Browne}},
  \bibinfo {author} {\bibfnamefont {S.}~\bibnamefont {Scheel}},\ and\ \bibinfo
  {author} {\bibfnamefont {M.}~\bibnamefont {Plenio}},\ }\bibfield  {title}
  {\bibinfo {title} {Distillation of continuous-variable entanglement with
  optical means},\ }\href {https://doi.org/10.1016/j.aop.2003.12.008}
  {\bibfield  {journal} {\bibinfo  {journal} {Ann. Phys.}\ }\textbf {\bibinfo
  {volume} {311}},\ \bibinfo {pages} {431} (\bibinfo {year}
  {2004})}\BibitemShut {NoStop}%
\bibitem [{\citenamefont {Campbell}\ and\ \citenamefont
  {Eisert}(2012)}]{Campbell2012}%
  \BibitemOpen
  \bibfield  {author} {\bibinfo {author} {\bibfnamefont {E.~T.}\ \bibnamefont
  {Campbell}}\ and\ \bibinfo {author} {\bibfnamefont {J.}~\bibnamefont
  {Eisert}},\ }\bibfield  {title} {\bibinfo {title} {Gaussification and
  entanglement distillation of continuous-variable systems: A unifying
  picture},\ }\href {https://doi.org/10.1103/physrevlett.108.020501} {\bibfield
   {journal} {\bibinfo  {journal} {Phys. Rev. Lett.}\ }\textbf {\bibinfo
  {volume} {108}},\ \bibinfo {pages} {020501} (\bibinfo {year}
  {2012})}\BibitemShut {NoStop}%
\bibitem [{\citenamefont {Seshadreesan}\ \emph {et~al.}(2019)\citenamefont
  {Seshadreesan}, \citenamefont {Krovi},\ and\ \citenamefont
  {Guha}}]{Seshadreesan2019}%
  \BibitemOpen
  \bibfield  {author} {\bibinfo {author} {\bibfnamefont {K.~P.}\ \bibnamefont
  {Seshadreesan}}, \bibinfo {author} {\bibfnamefont {H.}~\bibnamefont
  {Krovi}},\ and\ \bibinfo {author} {\bibfnamefont {S.}~\bibnamefont {Guha}},\
  }\bibfield  {title} {\bibinfo {title} {Continuous-variable entanglement
  distillation over a pure loss channel with multiple quantum scissors},\
  }\href {https://doi.org/10.1103/PhysRevA.100.022315} {\bibfield  {journal}
  {\bibinfo  {journal} {Phys. Rev. A}\ }\textbf {\bibinfo {volume} {100}},\
  \bibinfo {pages} {022315} (\bibinfo {year} {2019})}\BibitemShut {NoStop}%
\bibitem [{\citenamefont {Furrer}\ and\ \citenamefont
  {Munro}(2018)}]{Furrer2018}%
  \BibitemOpen
  \bibfield  {author} {\bibinfo {author} {\bibfnamefont {F.}~\bibnamefont
  {Furrer}}\ and\ \bibinfo {author} {\bibfnamefont {W.~J.}\ \bibnamefont
  {Munro}},\ }\bibfield  {title} {\bibinfo {title} {{Repeaters for
  continuous-variable quantum communication}},\ }\href
  {https://doi.org/10.1103/PhysRevA.98.032335} {\bibfield  {journal} {\bibinfo
  {journal} {Phys. Rev. A}\ }\textbf {\bibinfo {volume} {98}},\ \bibinfo
  {pages} {032335} (\bibinfo {year} {2018})}\BibitemShut {NoStop}%
\bibitem [{\citenamefont {Lund}\ and\ \citenamefont {Ralph}(2009)}]{Lund2009}%
  \BibitemOpen
  \bibfield  {author} {\bibinfo {author} {\bibfnamefont {A.~P.}\ \bibnamefont
  {Lund}}\ and\ \bibinfo {author} {\bibfnamefont {T.~C.}\ \bibnamefont
  {Ralph}},\ }\bibfield  {title} {\bibinfo {title} {{Continuous-variable
  entanglement distillation over a general lossy channel}},\ }\href
  {https://doi.org/10.1103/PhysRevA.80.032309} {\bibfield  {journal} {\bibinfo
  {journal} {Phys. Rev. A}\ }\textbf {\bibinfo {volume} {80}},\ \bibinfo
  {pages} {032309} (\bibinfo {year} {2009})}\BibitemShut {NoStop}%
\bibitem [{\citenamefont {Hu}\ \emph {et~al.}(2017)\citenamefont {Hu},
  \citenamefont {Liao},\ and\ \citenamefont {Zubairy}}]{Zubairy2017}%
  \BibitemOpen
  \bibfield  {author} {\bibinfo {author} {\bibfnamefont {L.}~\bibnamefont
  {Hu}}, \bibinfo {author} {\bibfnamefont {Z.}~\bibnamefont {Liao}},\ and\
  \bibinfo {author} {\bibfnamefont {M.~S.}\ \bibnamefont {Zubairy}},\
  }\bibfield  {title} {\bibinfo {title} {{Continuous-variable entanglement via
  multiphoton catalysis}},\ }\href {https://doi.org/10.1103/PhysRevA.95.012310}
  {\bibfield  {journal} {\bibinfo  {journal} {Phys. Rev. A}\ }\textbf {\bibinfo
  {volume} {95}},\ \bibinfo {pages} {012310} (\bibinfo {year}
  {2017})}\BibitemShut {NoStop}%
\bibitem [{\citenamefont {Birrittella}\ \emph {et~al.}(2018)\citenamefont
  {Birrittella}, \citenamefont {{El Baz}},\ and\ \citenamefont
  {Gerry}}]{Birrittella2018}%
  \BibitemOpen
  \bibfield  {author} {\bibinfo {author} {\bibfnamefont {R.~J.}\ \bibnamefont
  {Birrittella}}, \bibinfo {author} {\bibfnamefont {M.}~\bibnamefont {{El
  Baz}}},\ and\ \bibinfo {author} {\bibfnamefont {C.~C.}\ \bibnamefont
  {Gerry}},\ }\bibfield  {title} {\bibinfo {title} {{Photon catalysis and
  quantum state engineering}},\ }\href
  {https://doi.org/10.1364/josab.35.001514} {\bibfield  {journal} {\bibinfo
  {journal} {J. Opt. Soc. Am. B}\ }\textbf {\bibinfo {volume} {35}},\ \bibinfo
  {pages} {1514} (\bibinfo {year} {2018})}\BibitemShut {NoStop}%
\bibitem [{\citenamefont {Eaton}\ \emph {et~al.}()\citenamefont {Eaton},
  \citenamefont {Nehra},\ and\ \citenamefont {Pfister}}]{Eaton2019}%
  \BibitemOpen
  \bibfield  {author} {\bibinfo {author} {\bibfnamefont {M.}~\bibnamefont
  {Eaton}}, \bibinfo {author} {\bibfnamefont {R.}~\bibnamefont {Nehra}},\ and\
  \bibinfo {author} {\bibfnamefont {O.}~\bibnamefont {Pfister}},\ }\bibfield
  {title} {\bibinfo {title} {{Gottesman-Kitaev-Preskill state preparation by
  photon catalysis}},\ }\href {http://arxiv.org/abs/1903.01925} {\ }\Eprint
  {https://arxiv.org/abs/1903.01925} {arXiv:1903.01925 [quant-ph]} \BibitemShut
  {NoStop}%
\bibitem [{\citenamefont {Leonhardt}(1997)}]{Leonhardt1997}%
  \BibitemOpen
  \bibfield  {author} {\bibinfo {author} {\bibfnamefont {U.}~\bibnamefont
  {Leonhardt}},\ }\href
  {https://www.cambridge.org/ir/academic/subjects/physics/optics-optoelectronics-and-photonics/measuring-quantum-state-light?format=HB&isbn=9780521497305}
  {\emph {\bibinfo {title} {Measuring the quantum state of light}}}\ (\bibinfo
  {publisher} {Cambridge University Press},\ \bibinfo {address} {New York},\
  \bibinfo {year} {1997})\BibitemShut {NoStop}%
\bibitem [{\citenamefont {Vidal}\ and\ \citenamefont
  {Werner}(2002)}]{Vidal2002}%
  \BibitemOpen
  \bibfield  {author} {\bibinfo {author} {\bibfnamefont {G.}~\bibnamefont
  {Vidal}}\ and\ \bibinfo {author} {\bibfnamefont {R.~F.}\ \bibnamefont
  {Werner}},\ }\bibfield  {title} {\bibinfo {title} {{Computable measure of
  entanglement}},\ }\href {https://doi.org/10.1103/PhysRevA.65.032314}
  {\bibfield  {journal} {\bibinfo  {journal} {Phys. Rev. A}\ }\textbf {\bibinfo
  {volume} {65}},\ \bibinfo {pages} {032314} (\bibinfo {year}
  {2002})}\BibitemShut {NoStop}%
\bibitem [{\citenamefont {Plenio}(2005)}]{Plenio2005}%
  \BibitemOpen
  \bibfield  {author} {\bibinfo {author} {\bibfnamefont {M.~B.}\ \bibnamefont
  {Plenio}},\ }\bibfield  {title} {\bibinfo {title} {{Logarithmic Negativity: A
  Full Entanglement Monotone That is not Convex}},\ }\href
  {https://doi.org/10.1103/PhysRevLett.95.090503} {\bibfield  {journal}
  {\bibinfo  {journal} {Phys. Rev. Lett.}\ }\textbf {\bibinfo {volume} {95}},\
  \bibinfo {pages} {090503} (\bibinfo {year} {2005})}\BibitemShut {NoStop}%
\bibitem [{\citenamefont {Menicucci}\ \emph {et~al.}(2006)\citenamefont
  {Menicucci}, \citenamefont {{van Loock}}, \citenamefont {Gu}, \citenamefont
  {Weedbrook}, \citenamefont {Ralph},\ and\ \citenamefont
  {Nielsen}}]{Menicucci2006}%
  \BibitemOpen
  \bibfield  {author} {\bibinfo {author} {\bibfnamefont {N.~C.}\ \bibnamefont
  {Menicucci}}, \bibinfo {author} {\bibfnamefont {P.}~\bibnamefont {{van
  Loock}}}, \bibinfo {author} {\bibfnamefont {M.}~\bibnamefont {Gu}}, \bibinfo
  {author} {\bibfnamefont {C.}~\bibnamefont {Weedbrook}}, \bibinfo {author}
  {\bibfnamefont {T.~C.}\ \bibnamefont {Ralph}},\ and\ \bibinfo {author}
  {\bibfnamefont {M.~A.}\ \bibnamefont {Nielsen}},\ }\bibfield  {title}
  {\bibinfo {title} {Universal quantum computation with continuous-variable
  cluster states},\ }\href {https://doi.org/10.1103/PhysRevLett.97.110501}
  {\bibfield  {journal} {\bibinfo  {journal} {Phys. Rev. Lett.}\ }\textbf
  {\bibinfo {volume} {97}},\ \bibinfo {pages} {110501} (\bibinfo {year}
  {2006})}\BibitemShut {NoStop}%
\bibitem [{\citenamefont {Alexander}\ \emph {et~al.}(2018)\citenamefont
  {Alexander}, \citenamefont {Yokoyama}, \citenamefont {Furusawa},\ and\
  \citenamefont {Menicucci}}]{Alexander2018}%
  \BibitemOpen
  \bibfield  {author} {\bibinfo {author} {\bibfnamefont {R.~N.}\ \bibnamefont
  {Alexander}}, \bibinfo {author} {\bibfnamefont {S.}~\bibnamefont {Yokoyama}},
  \bibinfo {author} {\bibfnamefont {A.}~\bibnamefont {Furusawa}},\ and\
  \bibinfo {author} {\bibfnamefont {N.~C.}\ \bibnamefont {Menicucci}},\
  }\bibfield  {title} {\bibinfo {title} {Universal quantum computation with
  temporal-mode bilayer square lattices},\ }\href
  {https://doi.org/10.1103/PhysRevA.97.032302} {\bibfield  {journal} {\bibinfo
  {journal} {Phys. Rev. A}\ }\textbf {\bibinfo {volume} {97}},\ \bibinfo
  {pages} {032302} (\bibinfo {year} {2018})}\BibitemShut {NoStop}%
\bibitem [{\citenamefont {{Es'haqi-Sani}}\ \emph {et~al.}(2019)\citenamefont
  {{Es'haqi-Sani}}, \citenamefont {{Khazaei Nezhad}},\ and\ \citenamefont
  {Abdi}}]{Eshaqi2019}%
  \BibitemOpen
  \bibfield  {author} {\bibinfo {author} {\bibfnamefont {N.}~\bibnamefont
  {{Es'haqi-Sani}}}, \bibinfo {author} {\bibfnamefont {M.}~\bibnamefont
  {{Khazaei Nezhad}}},\ and\ \bibinfo {author} {\bibfnamefont {M.}~\bibnamefont
  {Abdi}},\ }\bibfield  {title} {\bibinfo {title} {Non-gaussian macroscopic
  entanglement of motion in a hybrid electromechanical device},\ }\href
  {https://doi.org/10.1103/PhysRevA.100.023845} {\bibfield  {journal} {\bibinfo
   {journal} {Phys. Rev. A}\ }\textbf {\bibinfo {volume} {100}},\ \bibinfo
  {pages} {023845} (\bibinfo {year} {2019})}\BibitemShut {NoStop}%
\bibitem [{\citenamefont {Genoni}\ \emph {et~al.}(2008)\citenamefont {Genoni},
  \citenamefont {Paris},\ and\ \citenamefont {Banaszek}}]{Genoni2008}%
  \BibitemOpen
  \bibfield  {author} {\bibinfo {author} {\bibfnamefont {M.~G.}\ \bibnamefont
  {Genoni}}, \bibinfo {author} {\bibfnamefont {M.~G.~A.}\ \bibnamefont
  {Paris}},\ and\ \bibinfo {author} {\bibfnamefont {K.}~\bibnamefont
  {Banaszek}},\ }\bibfield  {title} {\bibinfo {title} {{Quantifying the
  non-Gaussian character of a quantum state by quantum relative entropy}},\
  }\href {https://doi.org/10.1103/PhysRevA.78.060303} {\bibfield  {journal}
  {\bibinfo  {journal} {Phys. Rev. A}\ }\textbf {\bibinfo {volume} {78}},\
  \bibinfo {pages} {060303(R)} (\bibinfo {year} {2008})}\BibitemShut {NoStop}%
\bibitem [{\citenamefont {Genoni}\ and\ \citenamefont
  {Paris}(2010)}]{Genoni2010}%
  \BibitemOpen
  \bibfield  {author} {\bibinfo {author} {\bibfnamefont {M.~G.}\ \bibnamefont
  {Genoni}}\ and\ \bibinfo {author} {\bibfnamefont {M.~G.~A.}\ \bibnamefont
  {Paris}},\ }\bibfield  {title} {\bibinfo {title} {{Quantifying
  non-Gaussianity for quantum information}},\ }\href
  {https://doi.org/10.1103/PhysRevA.82.052341} {\bibfield  {journal} {\bibinfo
  {journal} {Phys. Rev. A}\ }\textbf {\bibinfo {volume} {82}},\ \bibinfo
  {pages} {052341} (\bibinfo {year} {2010})}\BibitemShut {NoStop}%
\bibitem [{\citenamefont {Marian}\ and\ \citenamefont
  {Marian}(2013)}]{Marian2013}%
  \BibitemOpen
  \bibfield  {author} {\bibinfo {author} {\bibfnamefont {P.}~\bibnamefont
  {Marian}}\ and\ \bibinfo {author} {\bibfnamefont {T.~A.}\ \bibnamefont
  {Marian}},\ }\bibfield  {title} {\bibinfo {title} {{Relative entropy is an
  exact measure of non-Gaussianity}},\ }\href
  {https://doi.org/10.1103/PhysRevA.88.012322} {\bibfield  {journal} {\bibinfo
  {journal} {Phys. Rev. A}\ }\textbf {\bibinfo {volume} {88}},\ \bibinfo
  {pages} {012322} (\bibinfo {year} {2013})}\BibitemShut {NoStop}%
\bibitem [{\citenamefont {Weedbrook}\ \emph {et~al.}(2012)\citenamefont
  {Weedbrook}, \citenamefont {Pirandola}, \citenamefont
  {Garc{\'{i}}a-Patr{\'{o}}n}, \citenamefont {Cerf}, \citenamefont {Ralph},
  \citenamefont {Shapiro},\ and\ \citenamefont {Lloyd}}]{Weedbrook2012}%
  \BibitemOpen
  \bibfield  {author} {\bibinfo {author} {\bibfnamefont {C.}~\bibnamefont
  {Weedbrook}}, \bibinfo {author} {\bibfnamefont {S.}~\bibnamefont
  {Pirandola}}, \bibinfo {author} {\bibfnamefont {R.}~\bibnamefont
  {Garc{\'{i}}a-Patr{\'{o}}n}}, \bibinfo {author} {\bibfnamefont {N.~J.}\
  \bibnamefont {Cerf}}, \bibinfo {author} {\bibfnamefont {T.~C.}\ \bibnamefont
  {Ralph}}, \bibinfo {author} {\bibfnamefont {J.~H.}\ \bibnamefont {Shapiro}},\
  and\ \bibinfo {author} {\bibfnamefont {S.}~\bibnamefont {Lloyd}},\ }\bibfield
   {title} {\bibinfo {title} {{Gaussian quantum information}},\ }\href
  {https://doi.org/10.1103/RevModPhys.84.621} {\bibfield  {journal} {\bibinfo
  {journal} {Rev. Mod. Phys.}\ }\textbf {\bibinfo {volume} {84}},\ \bibinfo
  {pages} {621} (\bibinfo {year} {2012})}\BibitemShut {NoStop}%
\bibitem [{\citenamefont {Barzanjeh}\ \emph {et~al.}(2012)\citenamefont
  {Barzanjeh}, \citenamefont {Abdi}, \citenamefont {Milburn}, \citenamefont
  {Tombesi},\ and\ \citenamefont {Vitali}}]{Barzanjeh2012}%
  \BibitemOpen
  \bibfield  {author} {\bibinfo {author} {\bibfnamefont {S.}~\bibnamefont
  {Barzanjeh}}, \bibinfo {author} {\bibfnamefont {M.}~\bibnamefont {Abdi}},
  \bibinfo {author} {\bibfnamefont {G.~J.}\ \bibnamefont {Milburn}}, \bibinfo
  {author} {\bibfnamefont {P.}~\bibnamefont {Tombesi}},\ and\ \bibinfo {author}
  {\bibfnamefont {D.}~\bibnamefont {Vitali}},\ }\bibfield  {title} {\bibinfo
  {title} {Reversible optical-to-microwave quantum interface},\ }\href
  {https://doi.org/10.1103/PhysRevLett.109.130503} {\bibfield  {journal}
  {\bibinfo  {journal} {Phys. Rev. Lett.}\ }\textbf {\bibinfo {volume} {109}},\
  \bibinfo {pages} {130503} (\bibinfo {year} {2012})}\BibitemShut {NoStop}%
\bibitem [{\citenamefont {Andrews}\ \emph {et~al.}(2014)\citenamefont
  {Andrews}, \citenamefont {Peterson}, \citenamefont {Purdy}, \citenamefont
  {Cicak}, \citenamefont {Simmonds}, \citenamefont {Regal},\ and\ \citenamefont
  {Lehnert}}]{Andrews2014}%
  \BibitemOpen
  \bibfield  {author} {\bibinfo {author} {\bibfnamefont {R.~W.}\ \bibnamefont
  {Andrews}}, \bibinfo {author} {\bibfnamefont {R.~W.}\ \bibnamefont
  {Peterson}}, \bibinfo {author} {\bibfnamefont {T.~P.}\ \bibnamefont {Purdy}},
  \bibinfo {author} {\bibfnamefont {K.}~\bibnamefont {Cicak}}, \bibinfo
  {author} {\bibfnamefont {R.~W.}\ \bibnamefont {Simmonds}}, \bibinfo {author}
  {\bibfnamefont {C.~A.}\ \bibnamefont {Regal}},\ and\ \bibinfo {author}
  {\bibfnamefont {K.~W.}\ \bibnamefont {Lehnert}},\ }\bibfield  {title}
  {\bibinfo {title} {Bidirectional and efficient conversion between microwave
  and optical light},\ }\href {https://doi.org/10.1038/nphys2911} {\bibfield
  {journal} {\bibinfo  {journal} {Nat. Phys.}\ }\textbf {\bibinfo {volume}
  {10}},\ \bibinfo {pages} {321} (\bibinfo {year} {2014})}\BibitemShut
  {NoStop}%
\bibitem [{\citenamefont {Palomaki}\ \emph {et~al.}(2013)\citenamefont
  {Palomaki}, \citenamefont {Teufel}, \citenamefont {Simmonds},\ and\
  \citenamefont {Lehnert}}]{Palomaki2013}%
  \BibitemOpen
  \bibfield  {author} {\bibinfo {author} {\bibfnamefont {T.~A.}\ \bibnamefont
  {Palomaki}}, \bibinfo {author} {\bibfnamefont {J.~D.}\ \bibnamefont
  {Teufel}}, \bibinfo {author} {\bibfnamefont {R.~W.}\ \bibnamefont
  {Simmonds}},\ and\ \bibinfo {author} {\bibfnamefont {K.~W.}\ \bibnamefont
  {Lehnert}},\ }\bibfield  {title} {\bibinfo {title} {Entangling mechanical
  motion with microwave fields},\ }\href
  {https://doi.org/10.1126/science.1244563} {\bibfield  {journal} {\bibinfo
  {journal} {Science}\ }\textbf {\bibinfo {volume} {342}},\ \bibinfo {pages}
  {710} (\bibinfo {year} {2013})}\BibitemShut {NoStop}%
\bibitem [{\citenamefont {Vitali}\ \emph {et~al.}(2007)\citenamefont {Vitali},
  \citenamefont {Gigan}, \citenamefont {Ferreira}, \citenamefont {Bohm},
  \citenamefont {Tombesi}, \citenamefont {Guerreiro}, \citenamefont {Vedral},
  \citenamefont {Zeilinger},\ and\ \citenamefont {Aspelmeyer}}]{Vitali2007}%
  \BibitemOpen
  \bibfield  {author} {\bibinfo {author} {\bibfnamefont {D.}~\bibnamefont
  {Vitali}}, \bibinfo {author} {\bibfnamefont {S.}~\bibnamefont {Gigan}},
  \bibinfo {author} {\bibfnamefont {A.}~\bibnamefont {Ferreira}}, \bibinfo
  {author} {\bibfnamefont {H.~R.}\ \bibnamefont {Bohm}}, \bibinfo {author}
  {\bibfnamefont {P.}~\bibnamefont {Tombesi}}, \bibinfo {author} {\bibfnamefont
  {A.}~\bibnamefont {Guerreiro}}, \bibinfo {author} {\bibfnamefont
  {V.}~\bibnamefont {Vedral}}, \bibinfo {author} {\bibfnamefont
  {A.}~\bibnamefont {Zeilinger}},\ and\ \bibinfo {author} {\bibfnamefont
  {M.}~\bibnamefont {Aspelmeyer}},\ }\bibfield  {title} {\bibinfo {title}
  {Optomechanical entanglement between a movable mirror and a cavity field},\
  }\href {https://doi.org/10.1103/PhysRevLett.98.030405} {\bibfield  {journal}
  {\bibinfo  {journal} {Phys. Rev. Lett.}\ }\textbf {\bibinfo {volume} {98}},\
  \bibinfo {pages} {030405} (\bibinfo {year} {2007})}\BibitemShut {NoStop}%
\bibitem [{\citenamefont {Genes}\ \emph {et~al.}(2008)\citenamefont {Genes},
  \citenamefont {Mari}, \citenamefont {Tombesi},\ and\ \citenamefont
  {Vitali}}]{Genes2008}%
  \BibitemOpen
  \bibfield  {author} {\bibinfo {author} {\bibfnamefont {C.}~\bibnamefont
  {Genes}}, \bibinfo {author} {\bibfnamefont {A.}~\bibnamefont {Mari}},
  \bibinfo {author} {\bibfnamefont {P.}~\bibnamefont {Tombesi}},\ and\ \bibinfo
  {author} {\bibfnamefont {D.}~\bibnamefont {Vitali}},\ }\bibfield  {title}
  {\bibinfo {title} {Robust entanglement of a micromechanical resonator with
  output optical fields},\ }\href {https://doi.org/10.1103/PhysRevA.78.032316}
  {\bibfield  {journal} {\bibinfo  {journal} {Phys. Rev. A}\ }\textbf {\bibinfo
  {volume} {78}},\ \bibinfo {pages} {032316} (\bibinfo {year}
  {2008})}\BibitemShut {NoStop}%
\bibitem [{\citenamefont {Abdi}\ and\ \citenamefont
  {Bahrampour}(2012)}]{Abdi2012a}%
  \BibitemOpen
  \bibfield  {author} {\bibinfo {author} {\bibfnamefont {M.}~\bibnamefont
  {Abdi}}\ and\ \bibinfo {author} {\bibfnamefont {A.~R.}\ \bibnamefont
  {Bahrampour}},\ }\bibfield  {title} {\bibinfo {title} {Improving the
  optomechanical entanglement and cooling by photothermal force},\ }\href
  {https://doi.org/10.1103/PhysRevA.85.063839} {\bibfield  {journal} {\bibinfo
  {journal} {Phys. Rev. A}\ }\textbf {\bibinfo {volume} {85}},\ \bibinfo
  {pages} {063839} (\bibinfo {year} {2012})}\BibitemShut {NoStop}%
\end{thebibliography}%

\end{document}